\documentclass[aps,preprint,groupedaddress, 11pt]{revtex4-1}

\usepackage{graphicx}
\graphicspath{{figures/}} 
\usepackage{latexsym}
\usepackage{amsmath,amsthm,amsfonts,amssymb,amscd}
\usepackage[mathscr]{eucal}
\usepackage{fullpage}
\usepackage{relsize}
\newcommand{\R}{\mathbb{R}}
\usepackage{enumerate}
\usepackage{fancyhdr}
\usepackage{mathrsfs}
\usepackage{xcolor}
\usepackage{listings}
\usepackage{hyperref}
\usepackage{esvect}
\usepackage{verbatim}
\usepackage{subcaption}
\usepackage[framed,numbered,autolinebreaks,useliterate]{mcode}
\usepackage[utf8]{inputenc}
\usepackage{listings}
\usepackage{xcolor} 

\usepackage{varwidth}

\hypersetup{%
  colorlinks=true,
  linkcolor=black,
  linkbordercolor={0 0 1}
}

\graphicspath{{Figures/}}

\newcommand{\eps}{\varepsilon}

\begin{document}


\title{Global Sensitivity Analysis of Plasma Instabilities via Active Subspaces}



\author{Soraya Terrab}
\email[]{sterrab@mines.edu}
\affiliation{Department of Applied Mathematics and Statistics\\ 
Colorado School of Mines\\
Golden, CO 80401, USA}

\author{Stephen Pankavich}
\email[]{pankavic@mines.edu}
\thanks{The authors were supported by National Science Foundation grants DMS-1551229 and DMS-1614586.}
\affiliation{Department of Applied Mathematics and Statistics\\ 
Colorado School of Mines\\
Golden, CO 80401, USA}


\date{\today}

\begin{abstract}
Active subspace analysis is a useful computational tool to identify and exploit the most important linear combinations in the
space of a model's input parameters.  
These directions depend inherently on a quantity of interest, which can be represented as a function from input parameters to model outputs.  
As the dynamics of many plasma models are driven by potentially uncertain parameter values, the utilization of active subspaces to perform global sensitivity analysis represents an important tool to understand how certain physical phenomena depend upon fluctuations in the values of these parameters.
In the current paper, we construct and implement new computational methods to quantify the induced uncertainty within the growth rate generated by perturbations in a collisionless plasma modeled by the one-dimensional Vlasov-Poisson system near an unstable, spatially-homogeneous equilibrium in the linear regime.

\end{abstract}

\pacs{}

\maketitle

\section{Introduction}
\label{Introduction}


A number of plasma models require the strict knowledge of input parameters in order to tune them to experimental studies or validate computational simulations.
For example, scientists studying Landau damping \cite{Landau, Ryutov}, i.e. the stability and exponential decay of a collective mode of oscillations in a plasma without the consideration of collisions among charged particles, must first identify the parameter regimes within which such behavior occurs in order to understand the inherent structure of solutions.
Another phenomenon of extreme interest within the plasma physics community is the study of plasma instabilities, such as the Two-Stream and Bump-on-Tail instabilities, which arise from perturbations in the particle distribution function near an unstable equilibrium.
These instabilities play a significant role in the theoretical study of plasmas and can be applied to better understand the properties of charged beams in particle accelerators and magnetic confinement devices \cite{Herr, OC}.

While prior studies \cite{BCP1, BMP, GPS, GPS2, GPS4, GPS5, Pankavich2020, Pankavich2021} focus on theoretical results concerning the behavior of a deterministic model for electrostatic plasma interactions, in reality, the physical system depends inherently upon knowledge of a variety of parameters, which are often themselves influenced or determined by experimental data. As such information may possess intrinsic uncertainty due to measurement errors, it is crucial to understand how this level of uncertainty can propagate within the output variables of the model. In the context of plasma instabilities \cite{AP}, tools from the fields of uncertainty quantification \cite{Aspaces, Salteli} and computational science \cite{IL} can allow scientists and engineers to make predictions regarding the influence of model parameters.
Hence, it is the central purpose of this article to quantify the uncertainty within the growth rate from an unstable equilibrium that is created by random fluctuations in model parameters, including the mean drift velocity and thermal velocity of an equilibrium distribution (denoted by $\mu$ and $\sigma^2$) and the amplitude and frequency of associated oscillations (i.e. $\alpha$ and $k$ herein).

\section{Plasma Instabilities}
\label{PI}
It is well known that collisionless plasmas can experience instabilities generated by perturbations from an associated equilibrium state.
For instance, such instabilities are known to occur in particle beams, i.e. accelerated streams of charged particles (ions and electrons). In the laboratory, these beams are created by particle accelerators, like cathode ray tubes and cyclotrons, while such phenomena are naturally created by strong electric fields, as occur in double layers.
Particle beams possess a wide range of applications to Inertial Confinement Fusion (ICF) \cite{DQ, Nuclear},
fast ignition fusion \cite{Mason},
astrophysics \cite{Alfven, Medvedev}, 
and high energy density physics \cite{Joshi},
among many others.
Here, the background plasma presents a means of current and charge neutralization
for charged particle beams, enabling the ballistic propagation of an intense beam pulse \cite{Tokluoglu}. 
However, the beam streaming through the background plasma can lead to the development of many different instabilities, including the Bump-on-Tail and Two-Stream instabilities, which we will investigate below. 

In order to capture this unstable behavior for an electrostatic plasma, we will utilize illustrative examples arising from the Vlasov-Poisson system posed in a spatially-periodic two-dimensional phase space $(x,v)$ with $0 \leq x \leq L$, $-\infty < v < \infty$, and $t \geq 0$ representing time. In dimensionless form, this model is given by
\begin{subequations}
\label{VPnd}
\begin{align}
\partial_t f + v \partial_x f - E \partial_v f = 0\\
\partial_x E = 1 - \int f(t,x,v) \ dv.
\end{align}
\end{subequations}
Here, $f(t,x,v)$ represents the electron distribution function with a fixed and normalized ionic background density, and $E(t,x)$ is the electric field induced by the charges in the system. This system of partial differential equations is obtained by reducing the dimension of the parameter space in the original (dimensional) model via rescaling (Appendix \ref{sec:AppA}).

In the case of linearized instability, i.e. the study of the Vlasov-Poisson system linearized about a spatially-homogeneous equilibrium state $f_{eq}(v)$, precise analytic results regarding the growth rate have been established~\cite{FC, Fowler, KagSyd}.  
Here, one assumes the perturbative solution 
$$f(t,x,v) = f_{eq}(v)\left (1 + \delta f(t,x,v) \right )$$ where the perturbation satisfies the plane wave form
$$\delta f(t,x,v) = \alpha \exp(i[kx -\omega t]),$$ while ignoring the contribution from the resulting nonlinear term in the Vlasov equation. In particular, for sufficiently small $\alpha$ with $0 < \alpha \ll 1$, the associated electric field grows exponentially in the time-asymptotic limit, and this rate of growth $\gamma$  can be completely determined for a given perturbation frequency $k$.  
More specifically, if we decompose the temporal frequency into its real and imaginary components
$$\omega(k) = \omega_R(k) + i \gamma(k),$$
where both $\omega_R$ and $\gamma$ are real-valued, then the induced electric field $E(t,x; k)$ grows exponentially with rate $\gamma(k)>0$ and oscillates at a frequency $\omega_R(k)$ where these values can be determined as roots of the dispersion function \cite{FC}
\begin{equation}
\label{dispersion}
\eps(k, \omega) = 1 - \frac{1}{k^2}\int \frac{f_{eq}'(v)}{v - \omega/k} \ dv.
\end{equation}
Namely, for a given wavenumber $k$, we define $\gamma(k) = \text{Im}(\omega(k))$
where $\eps(k, \omega(k)) = 0$.
Notice that such roots are independent of the perturbation amplitude $\alpha$.

In the current context, we will consider two specific equilibria - the so-called Two-Stream equilibrium and the Bi-Maxwellian (sometimes referred to as the ``double Maxwellian'') equilibrium - in order to perform numerical studies of the growth rate of perturbative solutions.
The Two-Stream equilibrium with mean drift velocity $\mu \in \mathbb{R}$ and average thermal velocity (or variance) $\sigma^2 > 0$ is given by the formula
$$f_{TS}(v) =\frac{1}{\sqrt{2\pi \sigma^2}} |v-\mu|^2 \exp \left (-\frac{1}{2\sigma^2} \vert v -\mu \vert^2 \right ). $$
Using the original dimensional variables (Appendix \ref{sec:AppB}), the velocity spread $\sigma^2$ can be represented as a constant multiple of $k_B T/m$
where $m$ is a particle mass, $T$ is temperature, and $k_B$ is the Boltzmann constant.

After a few standard calculations (Appendix \ref{sec:AppB}), one computes a more precise representation for the dispersion relation, namely
\begin{equation}
\label{epsTS}
\eps_{TS}(k,\omega) =1 - \frac{1}{k^2} \left[1 - 2A(u)^2 +2\left(A(u)-A(u)^3\right)Z(A(u))\right]
\end{equation}
where we have denoted the phase velocity by $u=\omega/k$, the function
$$A(u)= \frac{1}{\sqrt{2\sigma^2}}\left(u-\mu \right),$$ 
and $Z(\cdot)$ to be the plasma $Z$-function, given by
\begin{equation}
\label{Z}
Z(z) = \frac{1}{\sqrt{\pi}} \int_{-\infty}^\infty \frac{e^{-t^2}}{t-z} \ dt
\end{equation}
for any complex $z$.
Similar to analytic results concerning the Landau damping \cite{Landau, Penrose} of perturbations from the Maxwellian equilibrium, the roots of the dispersion function for the Two-Stream instability cannot be precisely obtained via analytic means and must be either estimated or computationally approximated in order to determine the corresponding growth rate $\gamma_{TS}(k, \mu, \sigma^2)$.

We will also study the behavior near a different equilibrium with similarly-defined parameters, namely the Bi-Maxwellian equilibrium function given by
$$f_{BM}(v) = \frac{\beta}{\sqrt{2\pi \sigma_1^2}} \exp \left (-\frac{1}{2\sigma_1^2} |v -\mu_1|^2 \right ) + \frac{1- \beta}{\sqrt{2\pi \sigma_2^2}} \exp \left (-\frac{1}{2\sigma_2^2} |v -\mu_2|^2 \right ), $$
along with the constraint $0 < \beta < 1$. 
Here, the equilibrium distribution is merely a convex combination of two Maxwellians, and the parameters $\mu_1, \mu_2$ and $\sigma_1^2, \sigma_2^2$ again represent the mean drift velocity and average thermal velocity of each separate Maxwellian, respectively, while the parameter $\beta$ controls the relative densities of the Maxwellians contributing to the velocity distribution.
In this case, the dispersion relation can be analogously simplified (Appendix \ref{sec:AppB}) to arrive at
\begin{equation}
\label{epsBM}
\eps_{BM}(k,\omega) = 1+\frac{\beta}{\sigma_1^2k^2}\left[1 + A_1(u)Z(A_1(u))\right] +\frac{1-\beta}{\sigma_2^2k^2}\left[1 + A_2(u)Z(A_2(u))\right],
\end{equation}
where $u$ and $Z$ are defined as before and 
$$A_i(u)= \frac{1}{\sqrt{2\sigma_i^2}}\left(u-\mu_i \right)$$
 for $i=1,2$.
As for the Two-Stream equilibrium, the zeros of the dispersion function cannot be analytically computed, and a computational approach is needed to determine the influence of parameter variations on the associated rate of instability, in this case given by $\gamma_{BM}(k, \mu_1, \mu_2, \sigma_1^2, \sigma_2^2, \beta)$.
Thus, we will utilize a global sensitivity analysis via active subspace methods described in the following section to investigate the relationship between the growth rate of the instability and specific parameters in the system.

\section{Active Subspaces and Global Sensitivity Analysis}
\label{active}

Modern simulations of plasma dynamics require several inputs, e.g., charges, masses, mean velocities, and temperatures, as well as, initial and boundary conditions, and then output a number of quantities of interest.
Though dimensional analysis and other reduction techniques can be used to reduce the size of the parameter space, plasma physicists and computational scientists often use these simulations to study the relationship between the original input parameters and subsequent output variables. 
The field of uncertainty quantification (UQ) generally aims to characterize quantities of interest in simulations subject to variability in the inputs. 
These characterizations often reduce to parameter studies that treat the inherent model as a mapping between inputs $p$ and a specified quantity of interest $g(p)$.
Such studies fall under the domain of sensitivity analysis, and several techniques -- such as parameter correlations \cite{SK}, Sobol indices \cite{IL}, and Morris screening \cite{Smith} -- exist that use a few simulation runs to screen the importance of input variables. 

Another approach is to identify important linear combinations of the inputs $p$ and focus parameter studies along these associated directions.
Active subspaces \cite{Aspaces, compAspaces} are defined by important directions in the high-dimensional space of inputs; once identified, scientists
can exploit the active subspace to enable otherwise infeasible parameter studies for expensive simulations \cite{Constantine1, Constantine2}.
More precisely, an active subspace is a low-dimensional linear subspace of the set of parameters, in which input perturbations along these directions alter the model's predictions more, on average, than perturbations which are orthogonal to the subspace. The identification of these subspaces allow for global, rather than local, sensitivity measurements of output variables with respect to parameters and often the construction of reduced-order models that greatly decrease the dimension of the input parameter space.
In the current context, they may allow one to answer particular physical questions, such as determining 
which parameters are most influential to the rate of instability or identifying the minimal and maximal values of this rate within a particular parameter regime. 

We begin with a description of the gradient-based active subspace method, which has been recently summarized within \cite{Aspaces}.
Let $p \in P = [-1,1]^m$ denote a vector of model inputs, where $m$ is a positive integer representing the number of parameters, and the space $P$ represents a \emph{normalized} set of parameter values.  Namely, we assume that the independent inputs have been shifted and scaled so that they are centered at the origin and possess unit variation.
Additionally, we assume the input space is equipped with a probability density function $\psi(p)$ that is strictly positive in the domain of the quantity of interest $g(p)$, zero outside the domain, and is normalized so that $\int_P \psi(p) \ dp = 1.$
In practice, $\psi$ identifies the set of input parameters of interest and quantifies their variability. 
Assume that $g: P \to \R$ is continuous, square-integrable with respect to the weight $\psi$, and differentiable with gradient vector $\nabla g \in \R^m$, which is also square-integrable with respect to $\psi$. 
The active subspace is then defined by the first $n < m$ eigenvectors of the $m \times m$ symmetric positive semi-definite matrix
\begin{equation}
\label{C}
C = \int_P \nabla g(p) \nabla g(p)^T \psi(p) \ dp =: W \Lambda W^T,
\end{equation}
where the right side of \eqref{C} represents the spectral decomposition of $C$ \cite{Axler, LVS}. In other words, $W$ represents the orthogonal matrix whose columns $w_\ell, \, (\ell = 1, \ldots, m)$ are the orthonormal eigenvectors of $C$, and $\Lambda$ is the diagonal matrix of eigenvalues of $C$, denoted $\lambda_1,...,\lambda_m$.
The matrix $C$ represents an average derivative functional which weights input values according to the probability density $\psi$.  
Additionally, we assume that the eigenvalues in $\Lambda$ (each of which must be non-negative) are listed in descending order and the associated eigenvectors are listed within the same column as their corresponding eigenvalues. 
The eigenvalue $\lambda_\ell$ measures the average change in $g$ subject to perturbations in $p$ along the corresponding eigenvector $w_\ell$, as they are related by the identity
\begin{equation}
\label{lambda_l}
\lambda_\ell = \int_P \vert \nabla g(p) \cdot w_\ell \vert^2 \psi(p) \ dp
\end{equation}
for $\ell = 1, ..., m$.

For example, if $\lambda_\ell = 0$, then $g$ is constant along the direction $w_\ell$, and directions along which $g$ is constant  can be ignored when studying the behavior of $g$ under changes in the parameter space $P$. 
Conversely, if the eigenvalue under consideration is large, then we may deduce from \eqref{lambda_l} that $g$ changes considerably in the direction of the corresponding eigenvector. 
Now, suppose that a spectral gap exists and the first $n < m$ eigenvalues are much larger than the trailing $m-n$. Let $W_1$ be the matrix containing the first $n$ columns of $W$. Then, as we will show below, a reasonable approximation for $g$ is
$g(p) \approx h(W_1^Tp)$,
where $h$ is the projection of $g$ onto the range of $W_1$, i.e. $h(y) = g(W_1y)$. 

Once the eigendecomposition involving $W$ and $\Lambda$ in \eqref{C} has been determined, the eigenvalues and eigenvectors can be  separated in the following way:
\begin{equation}
\label{mdecomp}
\Lambda = \begin{bmatrix} 
			\Lambda_1 & 0 \\
			0 & \Lambda_2 \\
		  \end{bmatrix} , \quad
W = \begin{bmatrix}
         W_1 & W_2 \\
	\end{bmatrix}
\end{equation}
where $\Lambda_1$ contains the ``large'' eigenvalues of $C$, $\Lambda_2$ contains the ``small'' eigenvalues, and $W_k$ contains the eigenvectors associated with each $\Lambda_k$, for $k=1, 2$. 
A simple way to differentiate between the ``large'' and ``small'' eigenvalues is to list them on a log plot from greatest to least and determine the appearance of a spectral gap. This gap will correspond to differences of at least an order of magnitude, and thus allow one to compartmentalize the greatest eigenvalues within $\Lambda_1$ and the remaining, lesser eigenvalues in $\Lambda_2$. A more systematic method of choosing the number of eigenvalues to store within $\Lambda_1$ can also be utilized, as developed in \cite{Aspaces}.

With the decomposition \eqref{mdecomp}, we can represent any element $p$ of the parameter space by
\begin{equation}
p = \underbrace{W W^T}_\text{$=I$} p = W_1 \underbrace{W_1^T p}_\text{$=q$} + W_2 \underbrace{W_2^T p}_\text{$=r$} = W_1 q + W_2 r.
\end{equation}
Thus, evaluating the quantity of interest at $p$ is equivalent to doing so at the point $W_1 q + W_2 r$, and we may approximate $g(p)$ using 
$$g(p) = g(W_1 q + W_2 r) \approx g(W_1q)  = g(W_1W_1^Tp) = h(W_1^Tp).$$
By the definition of $W_1$ and $W_2$, it's clear that small perturbations in $r$ will not, on average, alter the values of $g$. However, small perturbations in $q$ will, on average, change $g$ significantly. For this reason the outputs of $W_1$ are defined to be the \textit{active subspace} of the model and the outputs of $W_2$ are the corresponding \textit{inactive subspace}.  The linear combinations that generate these subspaces then represent the contributions of differing parameters in the model and describe the sensitivity of the quantity of interest with respect to parameter variations.

In general, the eigenvalues and eigenvectors of $C$ defined by \eqref{C} can be well-approximated computationally, using finite difference methods and Monte Carlo sampling. Though we only briefly outline the method below, full details can be found in \cite{Aspaces} (Algorithm 3.1) and \cite{compAspaces}. The numerical algorithm can be described concisely as follows:

\begin{enumerate} 
\item Draw $N$ parameter samples $\{ p_j \}_{j=1}^N$ independently according to the density $\psi$. 
\item For each parameter sample $p_j$, compute the gradient $\nabla_p g_j = \nabla_p g(p_j)$ for each entry $i=1,..,m$ by using the finite difference approximation $$\partial_{p_i} g(p_j) \approx \frac{g(p_j + \eps_i) - g(p_j)}{\vert \eps_i \vert}$$ where 
$(\eps_i)_k = \Delta \delta_{ik}$
represents a vector perturbation from the sampled parameter values, $\delta_{ik}$ is the Kronecker delta, and $\Delta>0$ can be taken as small as desired.
\item Approximate the matrix $C$ by the finite sum
\vspace{-20pt} 
\begin{center}
\begin{equation*}
C \approx \hat{C} = \frac{1}{N} \sum_{j = 1}^N  (\nabla_p g_j) (\nabla_p g_j)^T
\end{equation*}
\end{center}
\item Compute the corresponding eigendecomposition $\hat{C} = \hat{W}\hat{\Lambda}\hat{W}^T$.
\end{enumerate}
We note that the last step is equivalent to computing the Singular Value Decomposition (SVD) of the matrix
$$\frac{1}{\sqrt{N}} \left [ \nabla_p g_1 \ldots \nabla_p g_N \right ] = \hat{W} \sqrt{\hat{\Lambda}} \hat{V},$$
where it can be shown that the singular values are the square roots of the eigenvalues of $\hat{C}$ and the left singular vectors are the eigenvectors of $\hat{C}$. The SVD method of approximating $\hat{C}$ was developed first in \cite{Russi} and further utilized to study the global sensitivity of parameters within a variety of scientific models \cite{ConstantineDiaz, ConstantineDoostan, Diaz, PankLoudon}. 

\begin{table}[t]
\centering
\vspace{0.1in}
\begin{footnotesize}
\begin{tabular}{p {1.75cm} p {3.75cm} p {3.5cm}}
\hline \multicolumn{3}{c}{Two-Stream Instability} \\
\hline
Parameter & Quantity & Baseline \\
\hline
$\alpha$ & Perturbation size & N/A \\
$k$ & Perturbation frequency & $0.5$ \\
$\mu$ & Mean velocity & $0$ \\
$\sigma^2$ & Thermal velocity & $1$ \\
$p$ &  Parameter vector &  $[p_1, p_2, p_3] \propto [k, \mu, \sigma^2]$ \\
\hline\\
\end{tabular}
\end{footnotesize}
\caption{\footnotesize{Two-Stream Instability Parameters and their Baseline Values.}
}
\label{tab:paramvalues}
\vspace{-0.1in}
\end{table} 

Once $\hat{W}$ and $\hat{\Lambda}$ are computed, we further decompose the eigenspace into their active and inactive portions, namely $\hat{W}_1$ and $\hat{W}_2$, which correspond to the set of eigenvectors associated with the large eigenvalues along the diagonal of $\hat{\Lambda}_1$ and the small eigenvalues along the diagonal of $\hat{\Lambda}_2$, respectively. 
In practice and as we will find below, many systems possess a one-dimensional active subspace, so that $\hat{\Lambda}_1 \in \mathbb{R}$ and $\hat{W}_1 = w \in \mathbb{R}^m$. In such a scenario, the values of the vector $w$ represent the weights in a linear combination of the input parameters along which the quantity of interest is most variable. In this way, the entries of $w$ describe the relative importance of the parameters with respect to this quantity. For instance, if $w_2 \gg w_1$, then we generally expect $g(p)$ to vary more when the second entry of $p$ is altered from the $w$ direction than when the first entry of $p$ is altered. Similarly, if say $w_3 \approx 0$ then $g(p)$ does not change much on average when the third entry of $p$ is altered. With this information, the ultimate goal is to produce a model possessing reduced dimensional dependence, and this can be done using a \emph{sufficient summary plot}. In particular, if the active subspace is one-dimensional, then we have identified the single direction in the parameter space along which $g$ is most variable, and hence, the Monte Carlo sample points $\{ p_j \}_{j=1}^N$ can be used to construct an approximate model $h$ along this direction, given by $w^Tp$. To create the reduced model, a simple linear fit, or if greater precision is required a nonlinear least-squares curve fit, can be used. In the following section, we focus on using these tools in conjunction with numerical simulations of the dispersion function in order to generate nonlinear, low-dimensional models for the growth rate of plasma instabilities.

\begin{table}[t]
\centering
\vspace{0.1in}
\begin{footnotesize}
\begin{tabular}{p {1.25cm} p {3.75cm} p {2.5cm} | p {1.25cm} p {3.75cm} p {2.5cm}}
\hline \multicolumn{3}{c}{Double Beam} & \multicolumn{3}{c}{Bump-on-Tail}\\
\hline
Param. & Quantity & Baseline & Param. & Quantity & Baseline \\
\hline
$\alpha$ & Perturbation size & N/A & $\alpha$ & Perturbation size & N/A\\
$k$ & Perturbation frequency & $0.5$ & $k$ & Perturbation frequency & $0.5$\\
$\mu_1$, $\mu_2$ & Mean velocities & $0$, $4$ & $\mu_1$, $\mu_2$ & Mean velocities & $0$, $4$\\
$\sigma_1^2$, $\sigma_2^2$ & Thermal velocities & $0.5$, $0.5$  & $\sigma_1^2$, $\sigma_2^2$ & Thermal velocities & $0.5$, $0.5$  \\

$\beta$ & Scaling parameter & $0.5$ & $\beta$ & Scaling parameter & $0.8$ \\
& & & & & \\
\multicolumn{6}{c}{Parameter vector \qquad  $p = [p_1, p_2, p_3, p_4, p_5, p_6] \propto [k, \mu_1, \mu_2, \sigma_1^2, \sigma_2^2, \beta]$} \qquad \\
\hline\\
\end{tabular}
\end{footnotesize}
\caption{\footnotesize{Bi-Maxwellian Parameters and their Baseline Values.}
}
\label{tab:paramvaluesBiMax}
\vspace{-0.1in}
\end{table} 
\section{Computational Methods}
%
Now that the active subspace method has been described, we will demonstrate its utility for simulations of the instability rate by applying the aforementioned algorithm. Of course, prior to applying the procedure for computing a reduced model, we must first decide how to represent the original computational model; that is, how to approximate solutions of \eqref{VPnd}. In our simulations the quantity of interest is the instability growth rate $\gamma$, and this must be computed from a set of normalized input parameters, e.g. $p = [p_1, p_2, p_3] \propto [k, \mu, \sigma^2]$ for the Two-Stream equilibrium, where the symbol $\propto$ is merely used to abbreviate the linear relationship between the original variables $(k, \mu, \sigma^2)$ and their normalized counterparts $(p_1, p_2, p_3)$ given by \eqref{p} below. 
Hence, the computational model is expressed as a specific function $g$ so that $\gamma = g(p)$. For the remainder of the paper, the function $g(p)$ merely represents the process of approximating the exponential growth rate of the Vlasov-Poisson system \eqref{VPnd} by computing roots of the dispersion function associated with a specific equilibrium and its normalized parameters $p$. 
We note that the dispersion relation solver requires use of a numerical approximation of the plasma $Z$-function \cite{Zaghloul-Ali} to compute the temporal frequency $\omega$ that satisfies $\eps(k,\omega) = 0$ for a specified value of $k$. 
%
To be clear, instead of using the dispersion relation one can choose a nonlinear approximation method. For instance, any numerical method, including popular Particle-in-Cell \cite{BL}, operator-splitting \cite{ChengKnorr}, or Discontinuous Galerkin \cite{Gamba, Seal} methods, can be employed to simulate \eqref{VPnd} with a linear fit used to approximate $\gamma$ from the associated electric field values, as the active subspace approximation is independent of the chosen numerical model.
However, computing the roots of the dispersion relation is significantly faster and less expensive than any of these other methods, which is why we employ it here.

Notice that gradients of the growth rate are also required to implement the active subspace method. Since we do not have an explicit representation of this quantity of interest as a function of system parameters, we cannot explicitly compute the necessary gradients. Instead, we approximate them using the aforementioned finite difference scheme with a step size of $\Delta = 10^{-6}$. 
Thus, each parameter sample merely requires two roots of the dispersion relation to construct an approximate gradient of the growth rate mapping with respect to the normalized model parameters.
  
Additionally, in forthcoming simulations each sample is chosen so that every normalized parameter is uniformly distributed between $-1$ and $1$, i.e. $p_j \sim \mathcal{U}\left ([-1,1]^m \right)$ for $j = 1,...,N$, so that $\psi(p) = 2^{-m}$. 
In order to map this normalized parameter space onto the physically-relevant range of parameter values, we use the linear mapping
\begin{equation}
\label{p}
p_{\text{range}} = \frac{1}{2}\big (\mathrm{diag}(u - \ell) p_j + (u + \ell) \big ),
\end{equation}
for the random samples $\{p_j\}_{j=1}^N$,
where $u$ and $\ell$ are vectors containing the upper and lower bounds on the original parameters, respectively. Thus, the resulting vector $p_{\text{range}}$ represents the physical parameter values input to the model. 
For our purposes, the upper and lower limits, $u$ and $\ell$ are defined to incorporate a specific percentage (e.g., $1\%, 5\%, 25\%$) above and below the baseline values given in Tables~\ref{tab:paramvalues} and \ref{tab:paramvaluesBiMax}, respectively. For instance, in performing a $1\%$ perturbation from the baseline parameter vector $b$, we take $u = 1.01b$ and $\ell = 0.99b$. If a baseline value  is identically zero, as in the case of $\mu$, then for a $1\%$ perturbation we take $u = 0.01$ and $\ell=-0.01$.

In the next section, we provide results from simulation studies of the Two-Stream and Bi-Maxwellian distributions. 
All numerical simulations were conducted in MATLAB using a high-performance computing node with a 3.0 GHz Skylake 5118 24-Core processor and 192 GB of RAM. The code used to generate the results in this section is available at \url{https://github.com/sterrab/GSA_PlasmaInstabilities.git} \cite{GSA-Git}.
Average simulation times ranged from two to five seconds to generate $2^9$ trials.
The global sensitivity analysis provides an explicit representation of the growth rate $\gamma = g(p)$. As mentioned in the previous section, if a one-dimensional active subspace arises then such a multivariate function can be well-approximated as a scalar function $h(y)$, where $y = w^Tp$ and $w$ is the weight vector corresponding to the independent parameters. In this case, $w^Tp$ is merely a linear combination of the weighted parameters $p$. These parameter weights $w$ and functions $h(y)$ are presented for the various global sensitivity analyses along with the associated sufficient summary plots. 
%


\section{Computational Results}

The active subspace method was numerically implemented  for the linear parameter regime by computationally approximating roots of the dispersion relation $\eps (k,\omega) = 0$ for $\omega$ given the parameter value $k$ and defining the output quantity of interest to be $\gamma = \text{Im}(\omega)$.
As the dispersion relation further depends upon parameters from the equilibrium distribution, the resulting growth rate will also be a function of these quantities.
Throughout the simulation study, the chosen equilibrium parameters are merely representative of a specific physical environment, and we note that any selection of parameter values can be utilized within the global sensitivity analysis.

\begin{figure}
    \centering
    \includegraphics[scale=0.5]{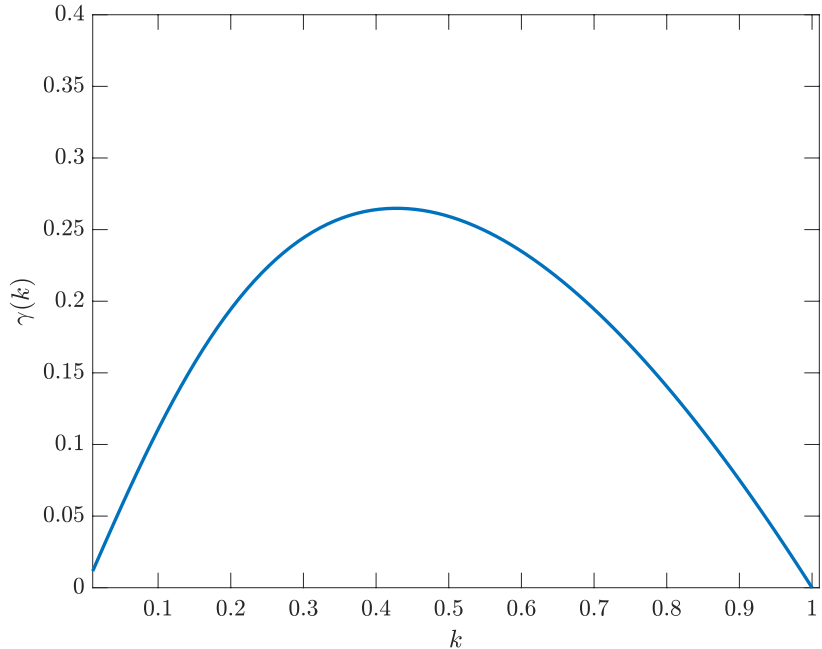}
    \caption{\footnotesize{\textit{Two-Stream Instability:} The growth rate as a function of $k$, indicating peak at $(k,\gamma) = (0.4241, 0.2649)$. The function matches that represented within \cite{Gamba}.}}
    \label{fig:V2Heath}
\end{figure}

In order to verify that our simulations match those of previous computational (but, not parameter) studies, we first performed a series of simulations for the Two-Stream instability with $\mu = 0$ and $\sigma^2 = 1$ and reproduced known results.
In particular, Fig.~\ref{fig:V2Heath} displays the resulting growth rate $\gamma$ as a function only of the wavenumber $k$ and results in outputs identical to that of \cite{Gamba}.

\subsection{Two-Stream Instability}
We begin by studying the behavior of unstable perturbations from the Two-Stream distribution, given by the equilibrium
$$f_{TS}(v) =\frac{1}{\sqrt{2\pi \sigma^2}} |v-\mu|^2 \exp \left (-\frac{1}{2\sigma^2} \vert v -\mu \vert^2 \right ). $$
Its associated dispersion relation (Appendix \ref{sec:AppB}) is 
$$\eps_{TS}(k,\omega) =1 - \frac{1}{k^2} \left[1 - 2A(u)^2 +2\left(A(u)-A(u)^3\right)Z(A(u))\right],$$
where $u=\omega/k$, $A(u)= \frac{1}{\sqrt{2\sigma^2}}\left(u-\mu \right)$, and $Z(\cdot)$ is the plasma Z-function defined by \eqref{Z}.
As the nonlinear growth of this instability has been shown numerically \cite{Gamba, Hou}, this three-parameter equilibrium serves as a useful test case prior to investigating more complex distributions.
Within these simulations, the parameters $k, \mu$, and $\sigma^2$ are shifted and scaled to create the normalized parameter vector $p=[p_1, p_2, p_3] \propto [k, \mu, \sigma^2]$, and their baseline values are set at $k=0.5, \mu=0$, and $\sigma^2 = 1$. The simulation study includes variations of 1\%, 15\%, 25\%, and 50\% of these nominal parameter values, and a total number of $512$ parameter samples were drawn in parallel. We note that taking, for instance, $\sigma^2 \in [0.5, 1.5]$ in the dimensionless framework is equivalent to considering perturbations of up to $50\%$ of the thermal velocity of the plasma.

\begin{figure}[t]
    \centering
    \includegraphics[scale=0.5]{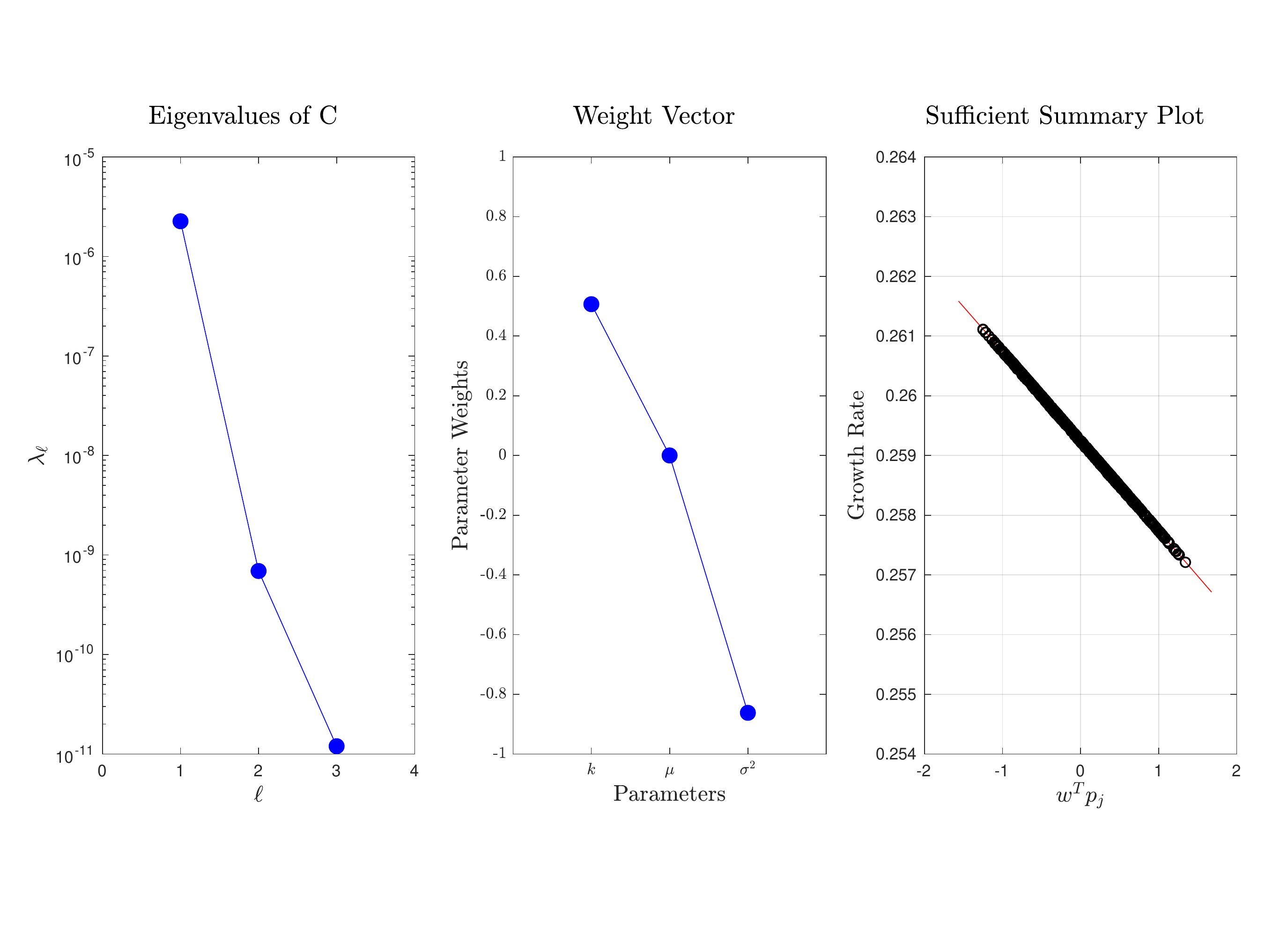}
    \vspace{-0.5in}
    \caption{\footnotesize{\textit{Two-Stream Instability:} Global Sensitivity Analysis (1\%) indicating the eigenvalues (left), parameter weights (center), and sufficient summary plot (right) of the one-dimensional active subspace, which represents $\eta_1 =99.97\%$ of the total variation.}}
    \label{fig:V2 1}
\end{figure}



\begin{figure}
    \centering
    \includegraphics[scale=0.5]{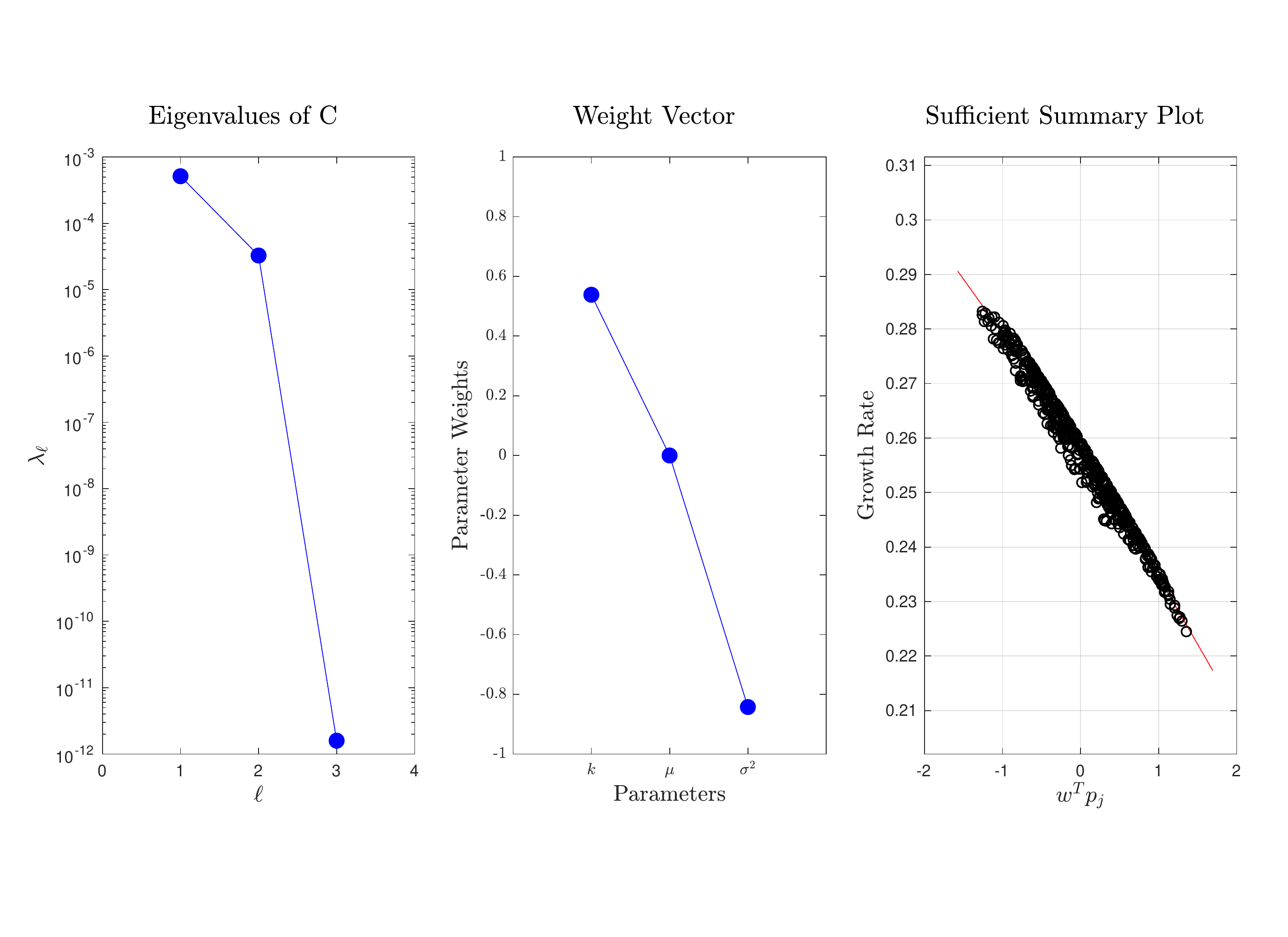}
    \vspace{-0.5in}
    \caption{\footnotesize{\textit{Two-Stream Instability:} Global Sensitivity Analysis (15\%) indicating the eigenvalues (left), parameter weights (center), and sufficient summary plot (right) of the one-dimensional active subspace, which represents $\eta_1 =94.00\%$ of the total variation.}}
    \label{fig:V2 15}
\end{figure}


\begin{figure}
    \centering
    \includegraphics[scale=0.5]{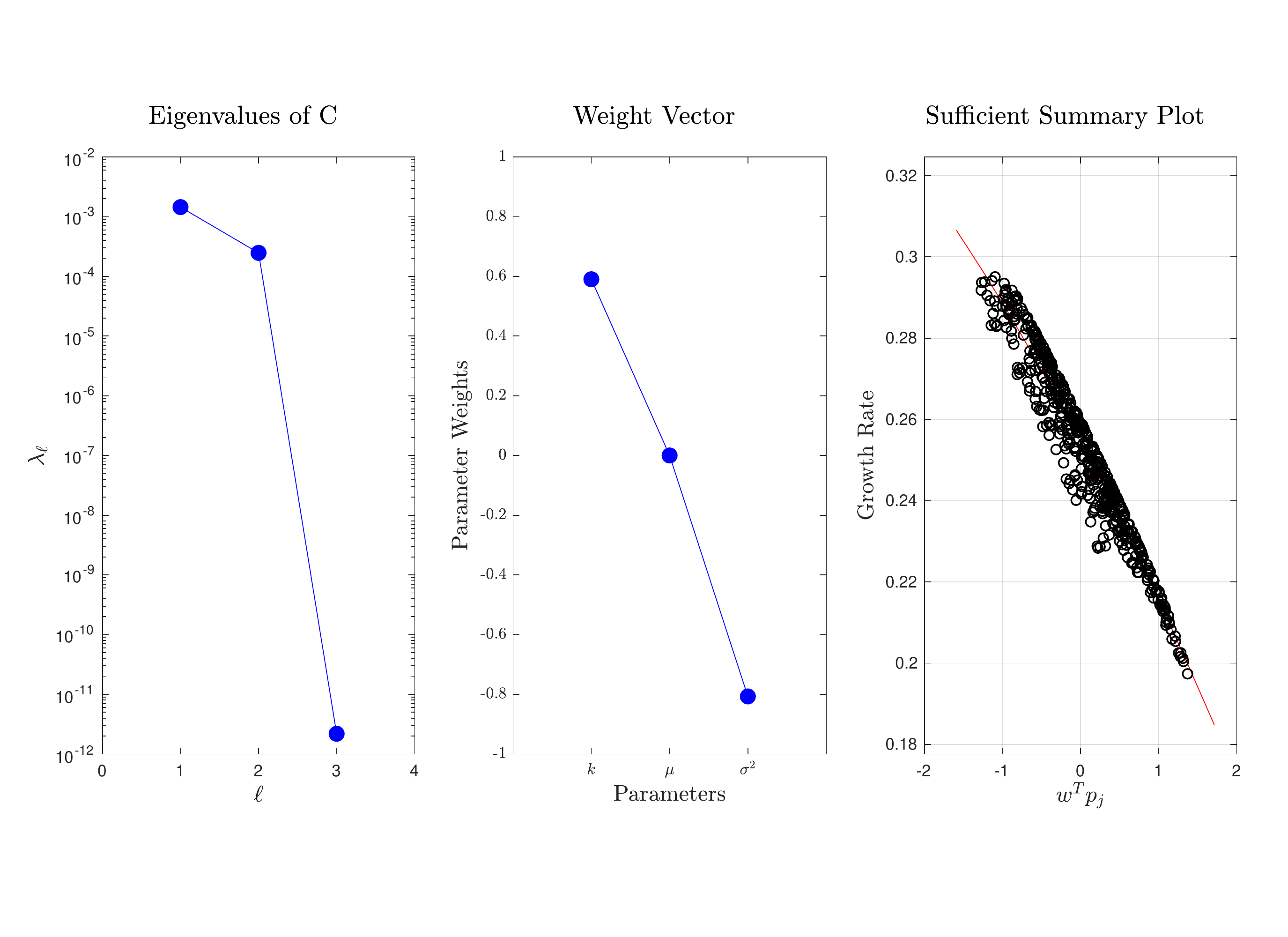}
    \vspace{-0.5in}
    \caption{\footnotesize{\textit{Two-Stream Instability:} Global Sensitivity Analysis (25\%) indicating the eigenvalues (left), parameter weights (center), and sufficient summary plot (right) of the one-dimensional active subspace, which represents $\eta_1 =85.38\%$ of the total variation.}}
    \label{fig:V2 25}
\end{figure}


With the simulation parameters complete, we first perform a $1\%$ perturbation simulation using the active subspace algorithm.
Figure \ref{fig:V2 1} displays three distinct panels of the resulting decomposition - the eigenvalues (listed in descending order) $\lambda_1 > ... > \lambda_4 =\ $diag$(\hat{\Lambda})$ of the covariance matrix $C$, the active subspace weight vector $w$, and a sufficient summary plot detailing the dependence of the output variable $\gamma \approx h(y)$ on the linear combination of input parameters $y = w^Tp$ with each of the $N$ data points representing a Monte Carlo sampling value $p_j$.

From the first panel, we can clearly identify a large spectral gap between the first and second eigenvalues. 
The second panel of the figure displays the eigenvector $w = [0.51, 1.2 \times 10^{-4}, -0.86]^T$ corresponding to the maximal eigenvalue $\lambda_1$, and it can be seen that two parameters possess associated weights greater than $0.5$, while the remaining parameter possesses a negligible weight. Hence, small perturbations in the former parameters, namely $k$ and $\sigma^2$, will significantly alter the value of $\gamma = g(p)$, as they are the most heavily weighted. Contrastingly, changes within the remaining parameter $\mu$, whose weight is near zero, will not have an appreciable affect on $\gamma$.
Intuitively, the effect on $\gamma$ of changing $\mu$ should be negligible as translations by $\mu$ in the velocity distribution function will not influence the growth rate. This idea can be generally inferred from \eqref{dispersion} in noticing that a change of variable $x = v -\mu$ in the dispersion relation integral will only alter $\text{Re}(\omega)$ and not $\text{Im}(\omega)$.

Finally, the third panel contains a plot of the quantity of interest $\gamma = g(p) \approx h(w^Tp)$. The horizontal axis contains values of the first active variable $y = w^Tp$, which represents a linear combination of the normalized parameters $p$ with weights given by entries of the first active variable vector $w$. This panel displays the linear functional form of the active subspace decomposition and shows a clear trend, namely that the growth rate $\gamma$ is a decreasing function of the active variable and can be well approximated by a linear function of $y$.
The linearity of this relationship is expected as a $1\%$ variation in parameters constitutes a small perturbation from their baseline values, and is similar to conducting a local sensitivity analysis near this point.
As the entries of the weight vector corresponding to $k$ and $\sigma^2$ are positive ($w_1> 0$) and negative ($w_3 < 0$), respectively, this further implies that the growth rate is decreasing in $k$ and increasing in $\sigma^2$.
Here, the growth rate is well-approximated as 
$$\gamma_{TS} \approx h \left (0.51 p_1 - 1.2\times 10^{-4} p_2 - 0.86 p_3 \right ),$$ where 
$$p_1 = 200 (k -0.5), \qquad
p_2 = 100 \mu, \qquad
p_3 = 100 (\sigma^2 -1)$$
are determined by inverting \eqref{p} for each parameter,
and $h$ is the linear function defined by 
$$h(y) = -0.0015y - 0.2592.$$
Simplifying these expressions, we can write an explicit representation for the growth rate in terms of the original parameters as
\begin{equation}
\label{gammaTS}
\gamma_{TS}(k,\mu, \sigma^2) \approx -0.0015 \left ( 102(k-0.5) - 0.012\mu - 86(\sigma^2 - 1) \right ) - 0.2592.
\end{equation}
Notice that this quantity is decreasing in the wavenumber and increasing with respect to the velocity spread. Thus, a less-focused pair of beams or reduced perturbation frequency will give rise to greater instability, while a more concentrated distribution or greater initial frequency will lessen the rate of instability. 

The results of additional simulations corresponding to 15\% and 25\% variations are displayed in Figs.~\ref{fig:V2 15}-\ref{fig:V2 25}. These plots include the eigenvalues on the left panel, the entries of the parameter weight vector $w$ of the first eigenvector in the center panel, and the sufficient summary plot of the one-dimensional active subspace in the right panel. Additionally, the reduced dimension approximation, calculated as 
$$\eta_1 = \frac{\lambda_1}{\sum_{i=1}^3 \lambda_i}$$ 
is included in the caption of each figure to demonstrate the percentage of the total variation captured by the one-dimensional subspace. 
This quantity is analogous to the eigenvalue ratio used in Principal Component Analysis (PCA). 
Hence, \eqref{gammaTS} can be used to compute the growth rate to over $99\%$ accuracy on the parameter space defined by a $1\%$ variation from baseline values.

\begin{table}[t]
    \centering
    \begin{tabular}{c| c | c | c}
     & \multicolumn{3}{c}{\textbf{Parameter Weights}}  \\
    Variation & \multicolumn{3}{c}{$w^Tp = w_1p_1 + w_2p_2 + w_3p_3$}  \\ 
     \cline{2-4}
     (\%) & $w_1$ & $w_2$ & $w_3$ \\
    \hline 
       1 & 0.5068 & $1.2054\times10^{-4}$ & -0.8620  \\
       5   &  0.5106 & $1.1535\times10^{-5}$ &  -0.8598 \\
      10    & 0.5225   & $ -2.1606\times10^{-6}$ & -0.8527  \\
      15    & 0.5385 & $2.7751\times10^{-6}$ & -0.8426  \\
      20    & 0.5605 & $-3.9831\times10^{-6}$ & -0.8281  \\
      25    & 0.5903 &$ 4.5345\times10^{-7}$ &  -0.8072  \\
      50    & 0.8974 & $-1.9074 \times 10^{-6}$ & -0.4412  \\
    \end{tabular}
    \caption{\footnotesize{Two-Stream Instability Parameter Weights with $p \propto [k,\mu,\sigma^2]$ . Notice that the weight $w_1$ associated to $k$ is increasing with the parameter variation from baseline values, while that of $\sigma^2$, namely $w_3$, decreases in magnitude. Hence, the wavenumber becomes more influential as the parameter space grows, and the thermal velocity less so.}}
    \label{tab:v2w}
\end{table}

As shown by the sufficient summary plot, the one-dimensional active subspace is essentially linear for the simulations presented in Figs.~\ref{fig:V2 1}-\ref{fig:V2 25}.  
The total variation captured by this approximation decreases as the parameter variations are increased, from $\eta_1 = 99.97\%$ for the 1\% variational study (Fig.~\ref{fig:V2 1})  to $\eta_1 = 85.38\%$ for the 25\% study (Fig.~\ref{fig:V2 25}). 

Table~\ref{tab:v2w} includes the parameter weights and corresponding $y=w^Tp$ scalar variable for the one-dimensional subspace. The exact polynomial fits are included in Table~\ref{tab:v2pf} and plotted in Fig.~\ref{fig:TS_Polyfits}. With larger variations, the coefficient of the quadratic term increases in magnitude from $-3.3\times 10^{-6}$ for the $1\%$ simulation to $-2.4 \times 10^{-3}$ for the $25\%$ simulation. This is expected as a first-order approximation is less informative away from baseline values. However, even within the 25\% simulation the linear term dominates as in Figs.~\ref{fig:V2 25} and \ref{fig:TS_Polyfits}.

Moreover, the spectral gap decreases as the variation percentage increases, displaying a loss in the amount of information that can be captured by using only a one-dimensional approximation along the direction of greatest output fluctuation.
Indeed, the disappearance of the spectral gap is visually documented from Fig.~\ref{fig:V2 15} to Fig.~\ref{fig:V2 25}. Here, the eigenvalues begin to cluster, and this greatly reduces the amount of variation contained within a one-dimensional active subspace decomposition. Additionally, $\eta_1 \approx 60\%$ for the $50\%$ variation from baseline values, so the information captured within a single active variable continues to decrease. Instead, a two-dimensional decomposition must be used to retain a suitable degree of variation in the values of the growth rate.

\begin{table}[t]
\hspace{-0.5in}
\begin{varwidth}[b]{0.42\linewidth}
    \centering
    \begin{tabular}{c|c |c|c }
      &  \multicolumn{3}{c}{\textbf{Polynomial Fit}}  \\ 
    Variation & \multicolumn{3}{c}{$h(w^Tp)=h(y)=a_2y^2+a_1y + a_0$} \\
    \cline{2-4}
     (\%) & $a_2$ & $a_1$ & $a_0$ \\
    \hline 
     1 &   $-3.2966\times 10^{-6}$ &  -0.0015 &    0.2592  \\
     5 &  -0.0001 &  -0.0075 &    0.2591 \\
     10 & -0.0003 &   -0.0150 &    0.2584  \\
     15 & -0.0008 &   -0.0223 &     0.2575  \\
     20 & -0.0014 &   -0.0297 &    0.2561  \\
     25 & -0.0024 &  -0.0366 &    0.2544  \\
     50 & -0.0313 &  -0.0556 &    0.2473  \\
    \end{tabular}
    \caption{\footnotesize{Two-Stream Instability Coefficients}}
    \label{tab:v2pf}
\end{varwidth}
    \hspace{0.2in}
\begin{minipage}[b]{0.42\linewidth}
    \includegraphics[width=90mm]{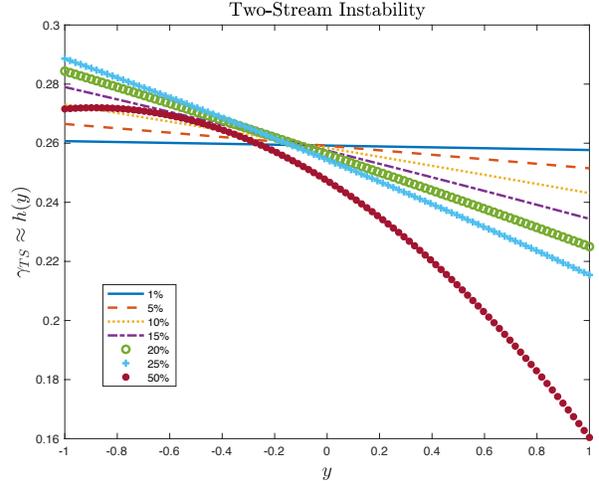}
    \vspace{-0.5in}
    \captionof{figure}{\footnotesize{Two-Stream Instability 2nd-order Polynomial Fits}}
    \label{fig:TS_Polyfits}
  \end{minipage}
\end{table}

\begin{figure}[t!]
\centering
\begin{subfigure}{.5\textwidth}
    \includegraphics[width=0.95\linewidth]{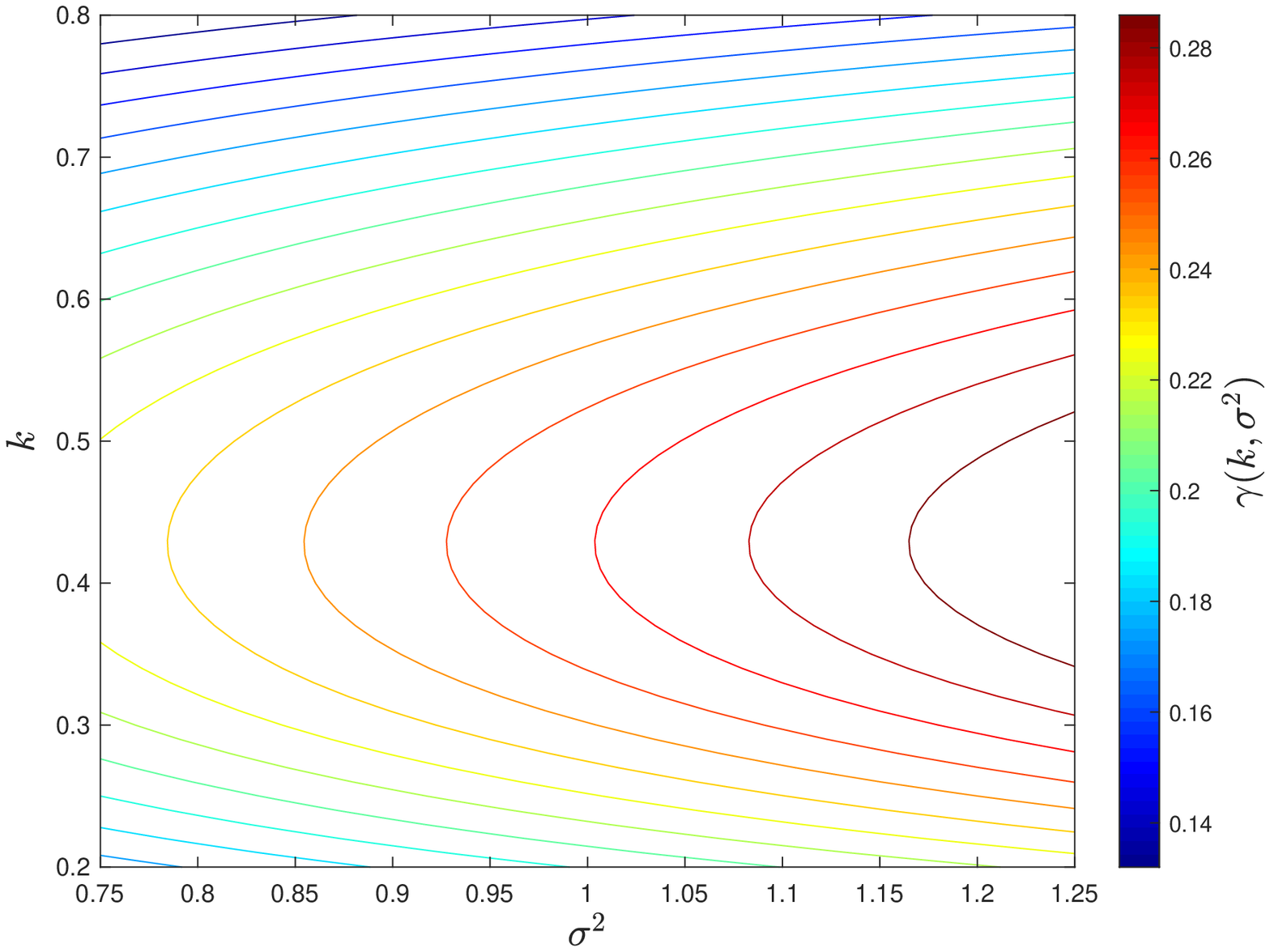}
\end{subfigure}%
\begin{subfigure}{.5\textwidth}
  \includegraphics[width=0.95\linewidth]{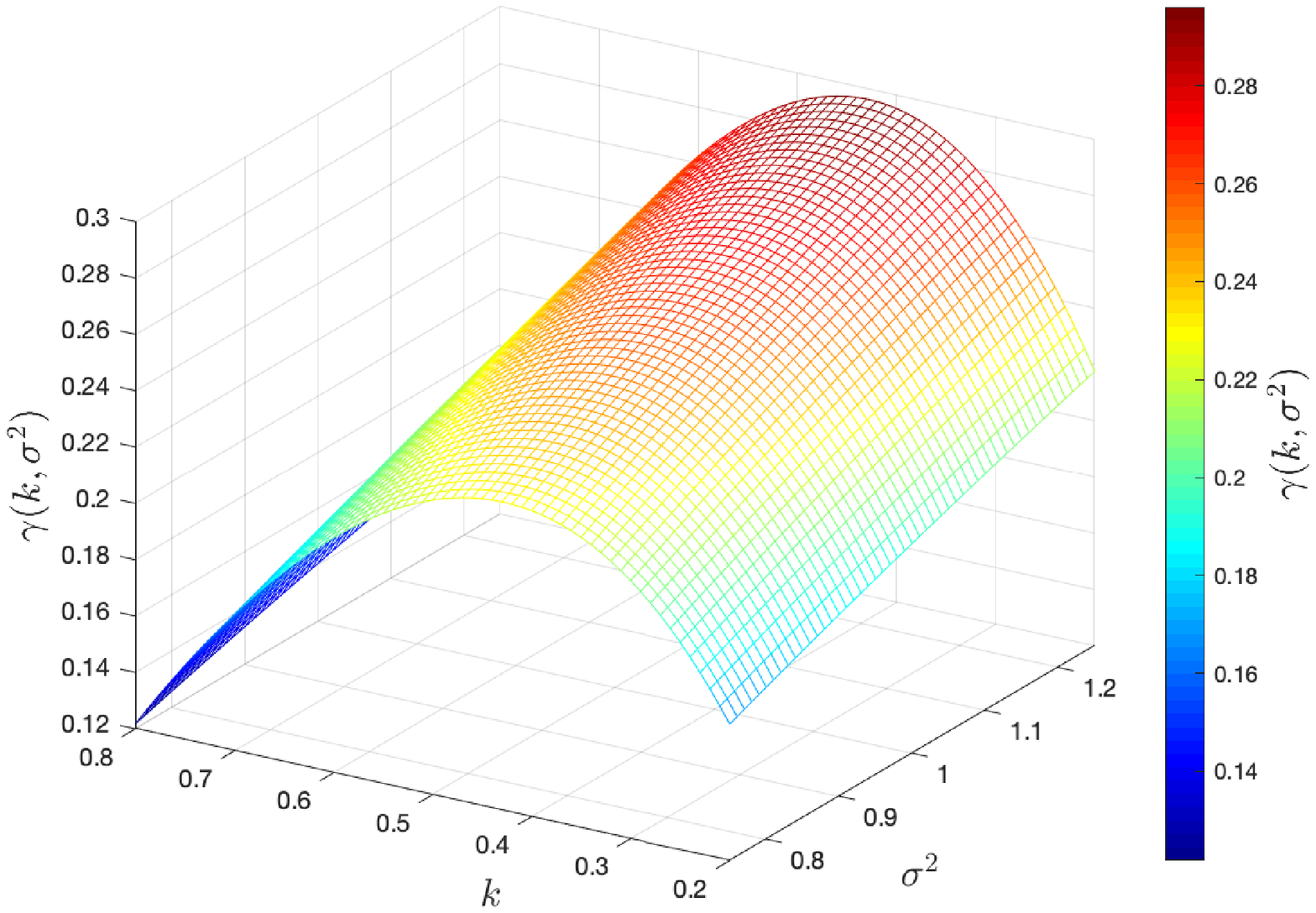}
\end{subfigure}
\caption{\footnotesize{Level curves (left) and three-dimensional plot (right) of the growth rate $\gamma(k,\sigma^2)$ as a function of the wavenumber and equilibrium variance. Each parameter is varied by $25\%$ of its baseline value.}}
\label{fig:V2twoparam}
\end{figure}

Alternatively, because the growth rate is insensitive to changes in the mean parameter $\mu$, we can more easily visualize it as a function of the remaining two parameters of the equilibrium distribution. In Fig.~\ref{fig:V2twoparam} we display a plot of the growth rate $\gamma(k, \sigma^2)$ as a function of the wavenumber and variance parameters with $\mu = 0$ fixed. Additionally, the level curves are displayed and the approximation of the weight vector $w$ is orthogonal to these curves near their baseline values.
For each of the parameter variations, the active subspace method allows for the construction of an inexpensive computational approximation and an explicit analytic formula to well approximate the growth rate as a function of model parameters.

\subsection{Bi-Maxwellian Distribution}
Next, we study the resulting behavior of unstable perturbations from the Bi-Maxwellian equilibrium distribution, given by
$$f_{eq}(v) = \frac{\beta}{\sqrt{2\pi \sigma_1^2}} \exp \left (-\frac{1}{2\sigma_1^2} (v -\mu_1)^2 \right ) + \frac{1-\beta}{\sqrt{2\pi \sigma_2^2}} \exp \left (-\frac{1}{2\sigma_2^2} (v -\mu_2)^2 \right ), $$
with the constraint $0 < \beta < 1$. The associated dispersion relation for this equilibrium (Appendix \ref{sec:AppB}) is given by 
\[ \eps_{BM}(k,\omega) = 1+\frac{\beta}{\sigma_1^2k^2}\left[1 + A_1(u)Z(A_1(u))\right] +\frac{1-\beta}{\sigma_2^2k^2}\left[1 + A_2(u)Z(A_2(u))\right],  \]
where $u=\omega/k$, 
$$A_i(u)= \frac{1}{\sqrt{2\sigma_i^2}}\left(u-\mu \right)$$
 for $i=1,2$, and $Z(\cdot)$ is again defined by \eqref{Z}.

\begin{figure}[t!]
\centering
\begin{subfigure}{.33\textwidth}
    \includegraphics[width=0.95\linewidth]{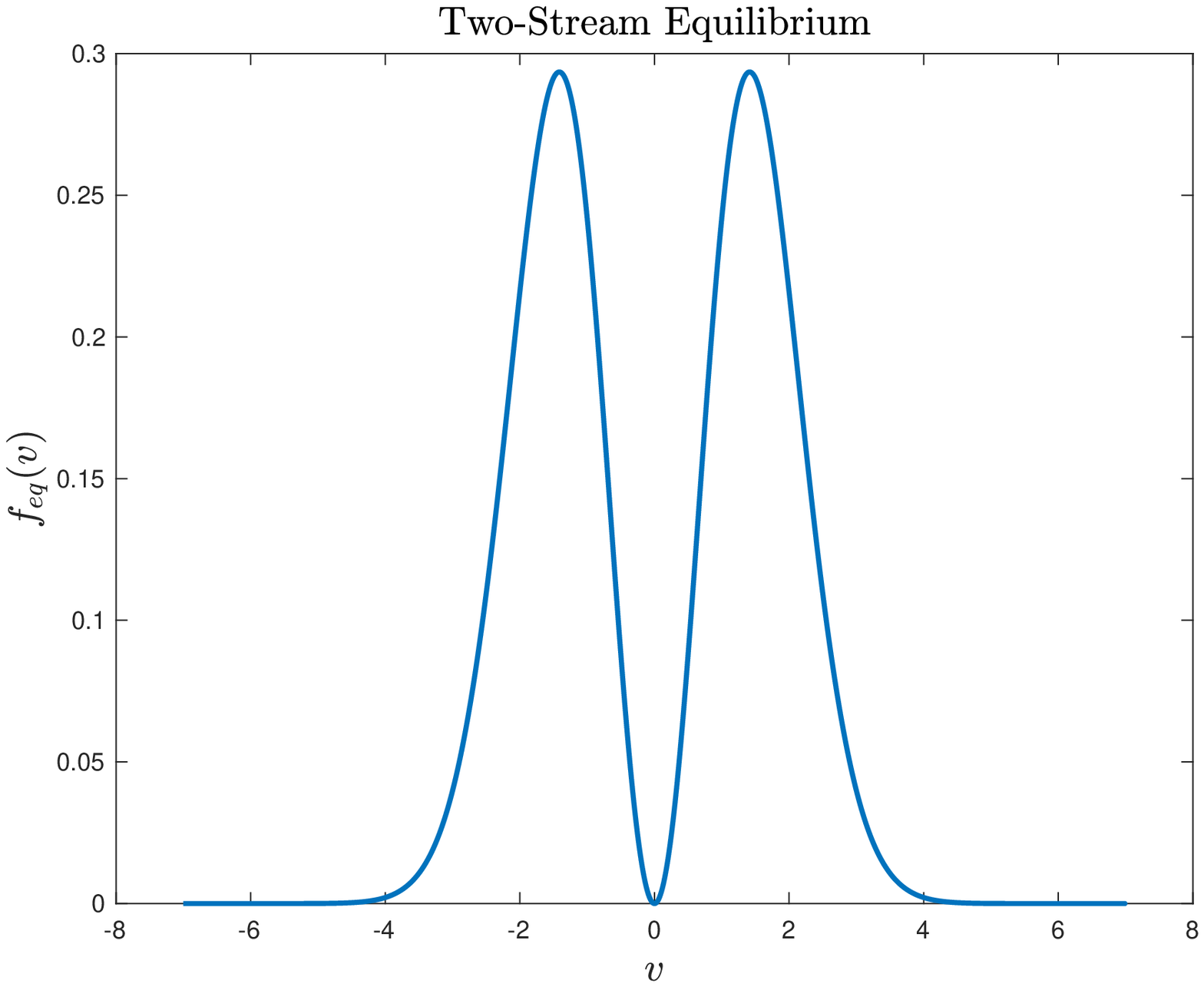}
\end{subfigure}%
\begin{subfigure}{.33\textwidth}
    \includegraphics[width=0.95\linewidth]{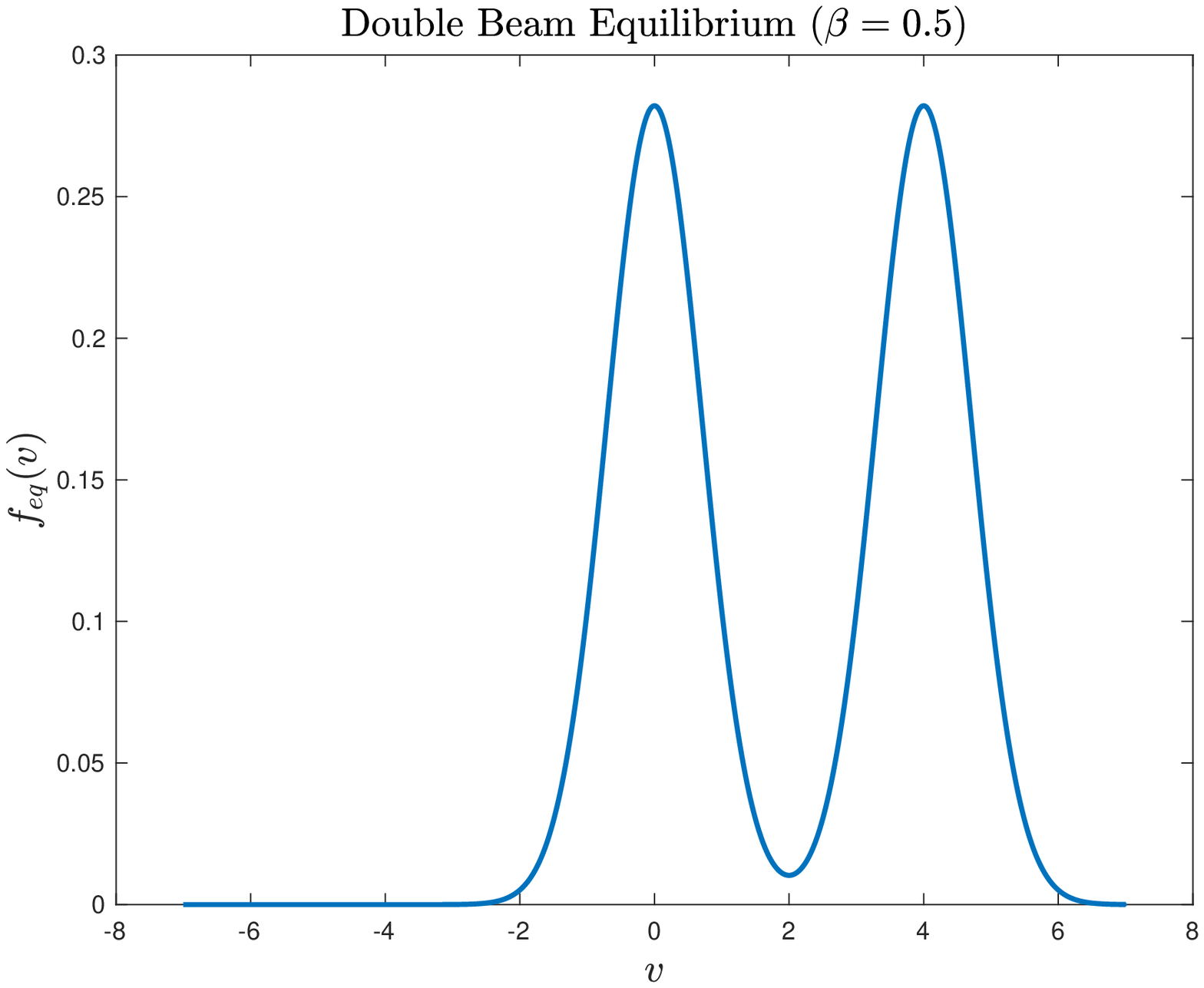}
\end{subfigure}%
\begin{subfigure}{.33\textwidth}
  \includegraphics[width=0.95\linewidth]{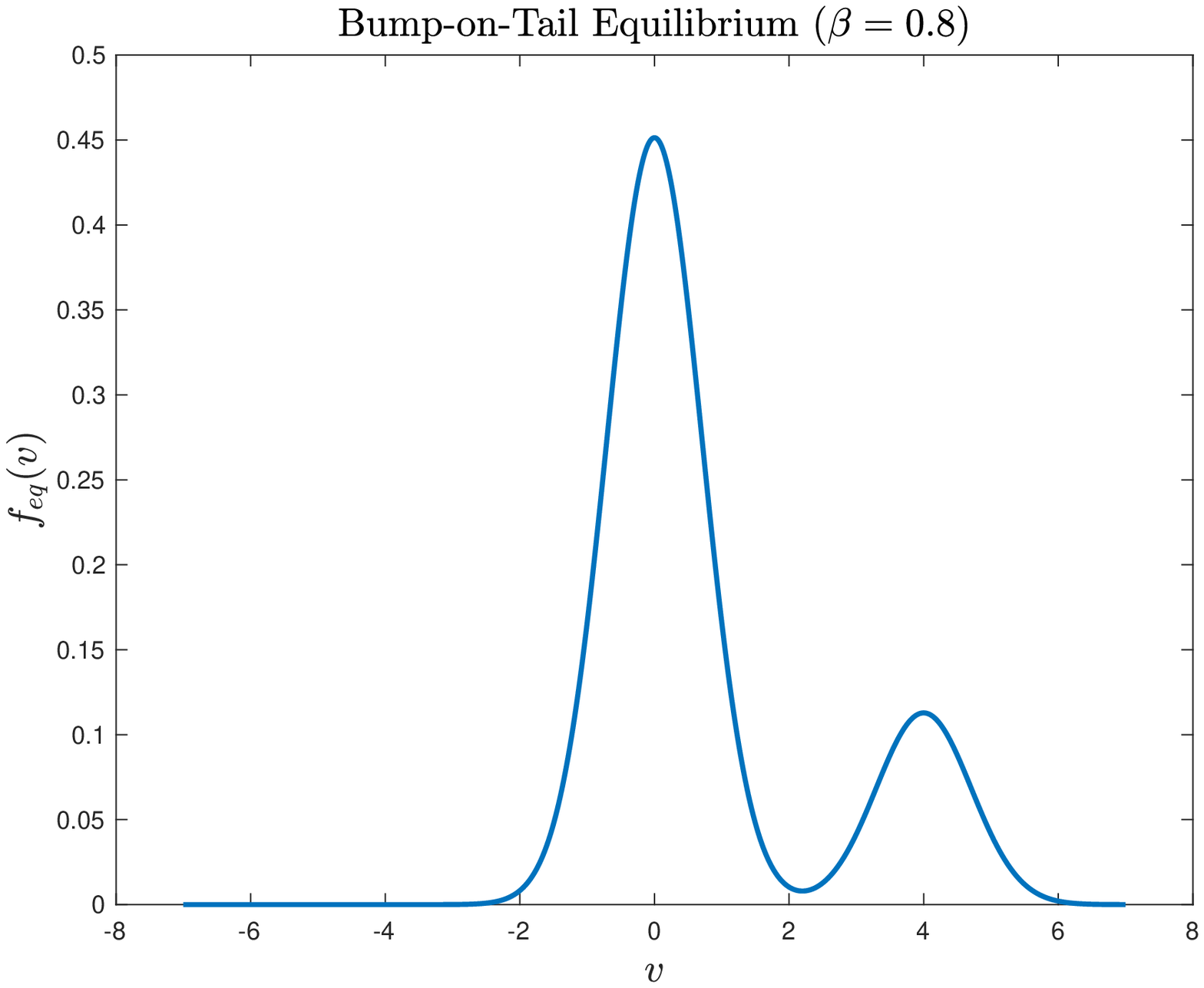}
\end{subfigure}
\caption{\footnotesize{Two-Stream equilibrium (left) and Bi-Maxwellian equilibria for $\beta = 0.5$  (center) and $\beta = 0.8$ (right).}} 
\label{fig:BiMax_plots}
\end{figure}

As before, the parameters (now $k, \mu_1, \mu_2, \sigma_1^2, \sigma_2^2$, and $\beta$) are shifted and scaled to create the normalized parameter vector $p=[p_1, p_2, p_3, p_4, p_5, p_6] \propto [k, \mu_1, \mu_2, \sigma_1^2, \sigma_2^2, \beta]$, and we will consider two different sets of baseline values - an instability driven by two distinct plasma beams with differing mean velocities but similar spread and strength that we will call the Double Beam Instability and the well-known Bump-on-Tail instability. The qualitative differences between these equilibria are displayed in Fig.~\ref{fig:BiMax_plots}.
Finally, to close this section we will construct an approximation for the Bi-Maxwellian growth rate that is global throughout the parameter space.

\subsubsection{Double Beam Instability}
In the first case, we investigate the behavior of the growth rate for a double-humped equilibrium that represents two ionic beams with similar densities and thermal velocities, but different mean velocities. Here, the baseline parameter values are $k=0.5, \mu_1=0, \mu_2=4, \sigma_1^2=0.5, \sigma_2^2=0.5$, and $\beta=0.5$. As in the previous section, a total number of $N=512$ samples were drawn in the parameter space.

\begin{figure}
    \centering
    \includegraphics[scale=0.5]{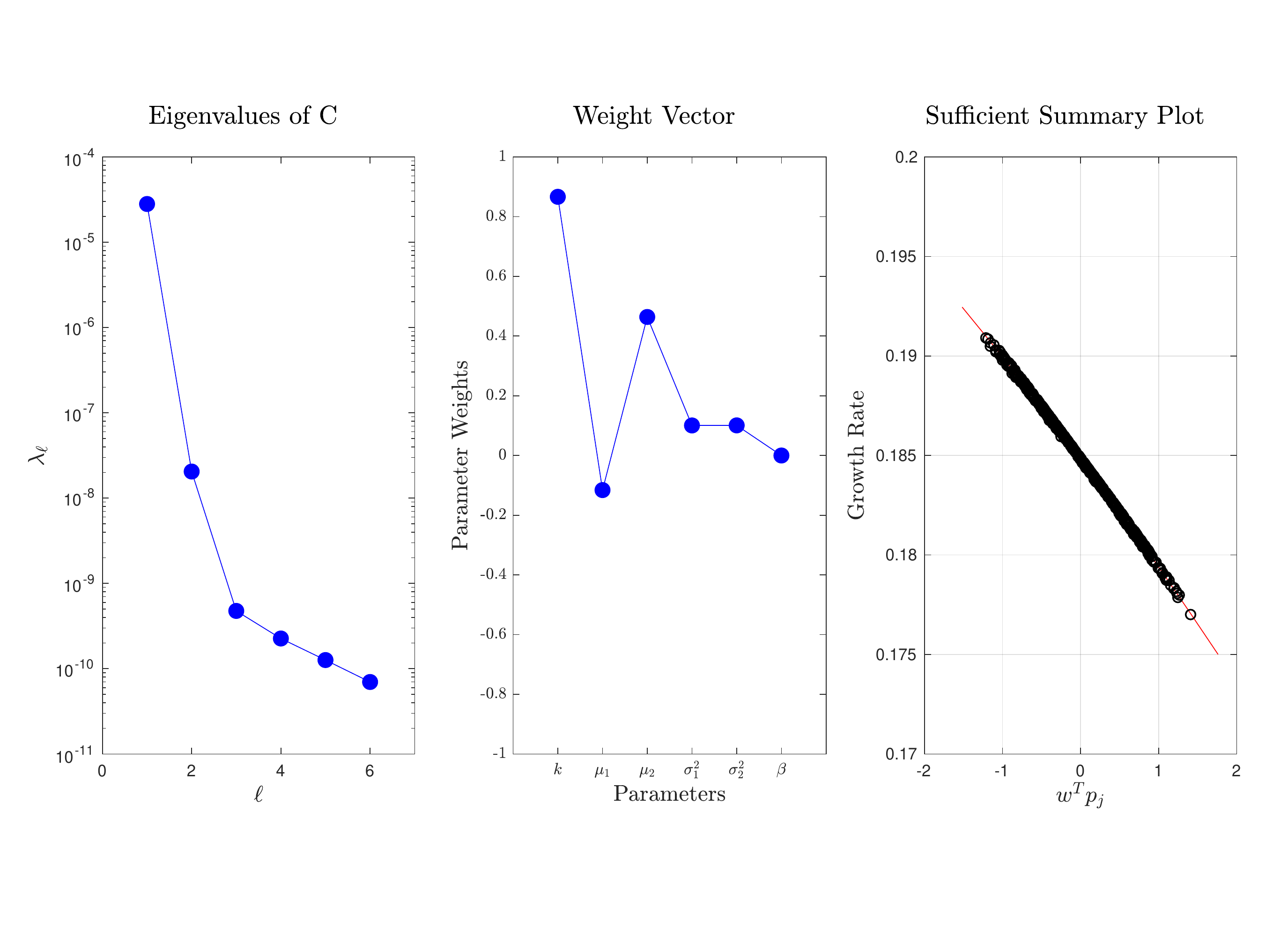}
    \vspace{-0.5in}
    \caption{\footnotesize{\textit{Double Beam Instability:} Global Sensitivity Analysis (1\%) indicating the eigenvalues (left), parameter weights (center), and sufficient summary plot (right) of the one-dimensional active subspace, which represents $\eta_1 = 99.92\%$ of the total variation.}}
    \label{fig:BiMax1}
\end{figure}

\begin{figure}
    \centering
    \includegraphics[scale=0.5]{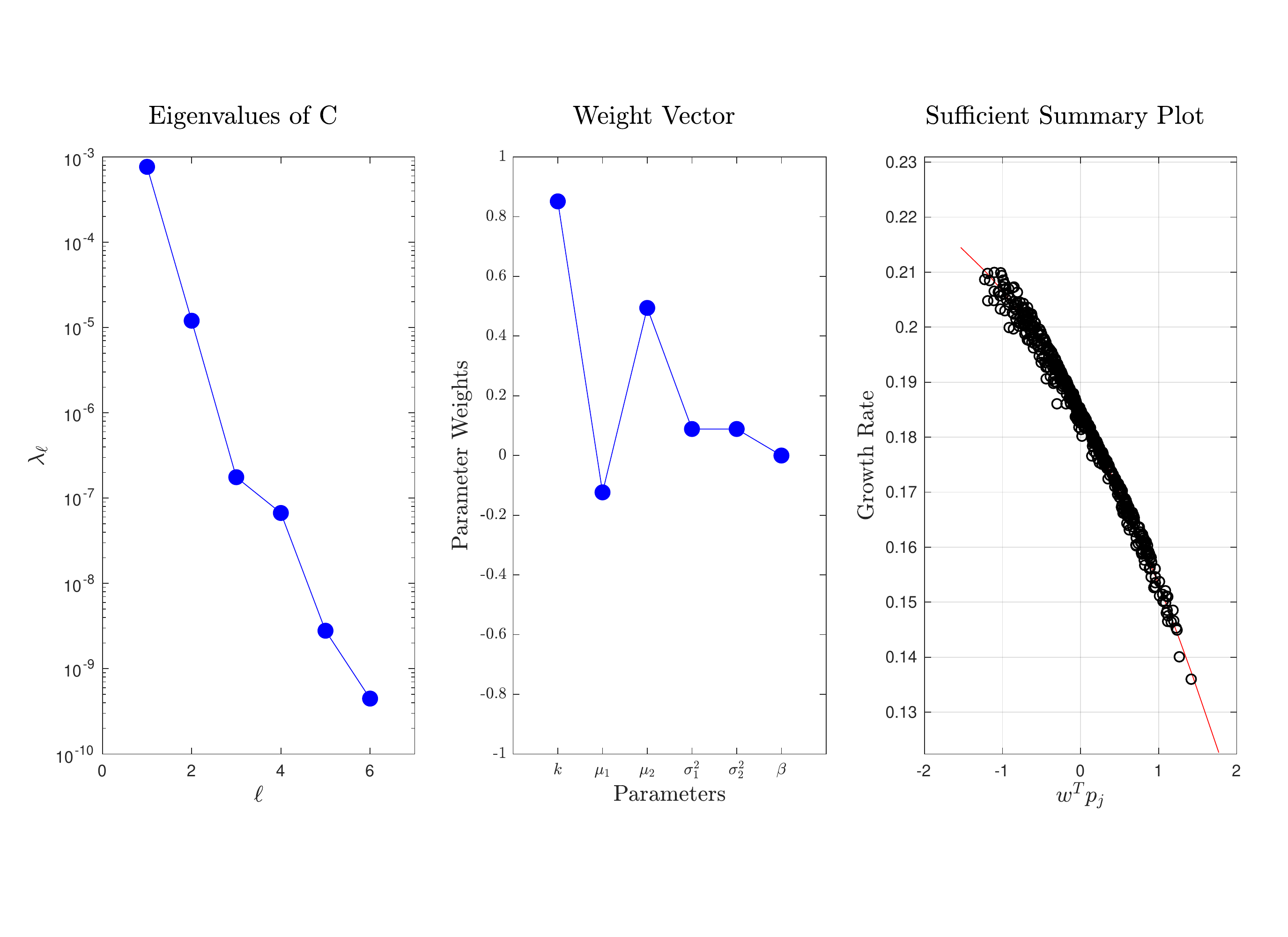}
    \vspace{-0.5in}
    \caption{\footnotesize{\textit{Double Beam Instability:} Global Sensitivity Analysis (5\%) indicating the eigenvalues (left), parameter weights (center), and sufficient summary plot (right) of the one-dimensional active subspace, which represents $\eta_1 = 98.42\%$ of the total variation.}}
    \label{fig:BiMax5}
\end{figure}

\begin{figure}
    \centering
    \includegraphics[scale=0.5]{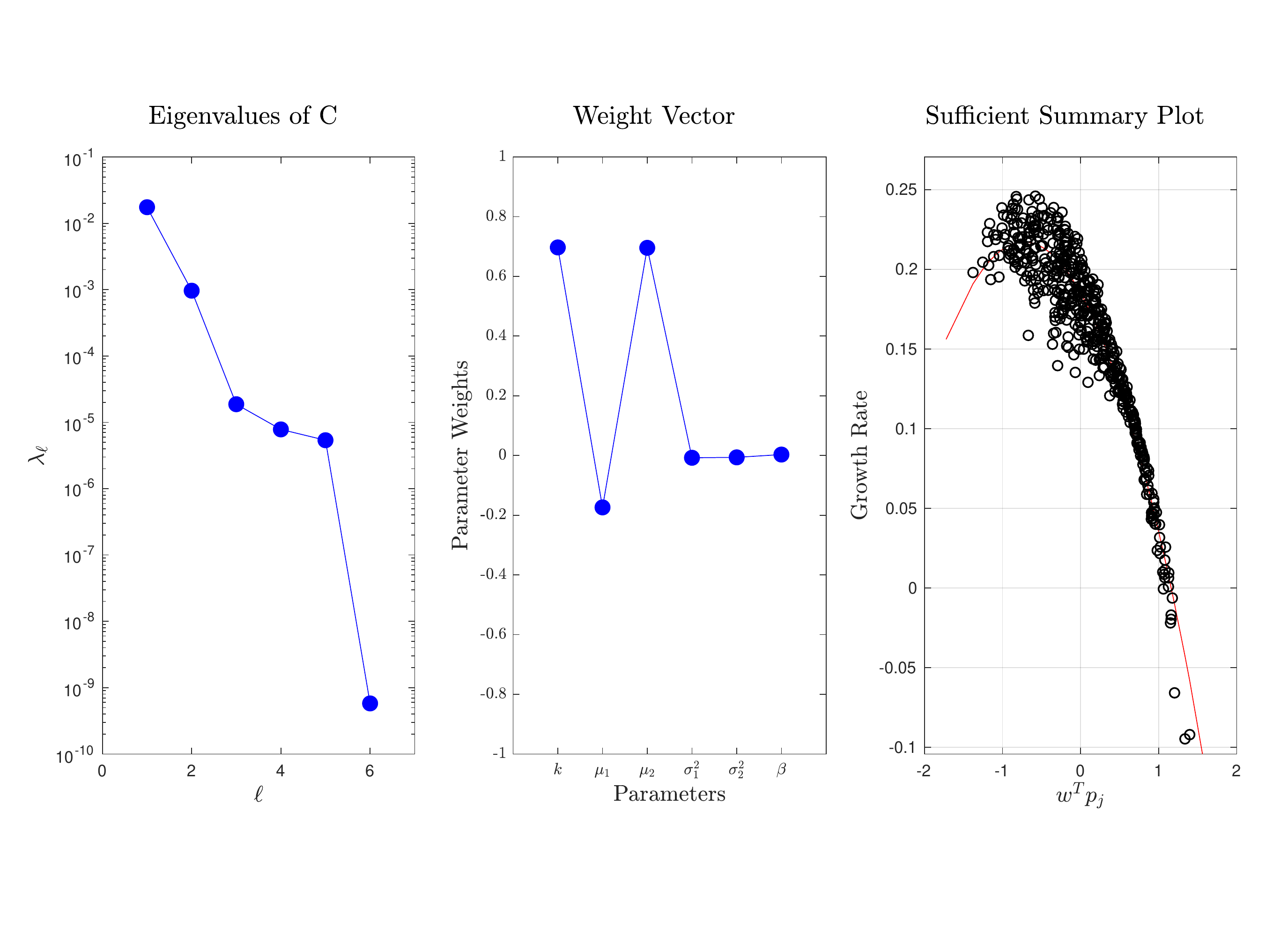}
    \vspace{-0.5in}
    \caption{\footnotesize{\textit{Double Beam Instability:} Global Sensitivity Analysis (15\%) indicating the eigenvalues (left), parameter weights (center), and sufficient summary plot (right) of the one-dimensional active subspace, which represents $\eta_1 = 94.61\%$ of the total variation.}}
    \label{fig:BiMax15}
\end{figure}

\begin{figure}
    \centering
    \includegraphics[scale=0.5]{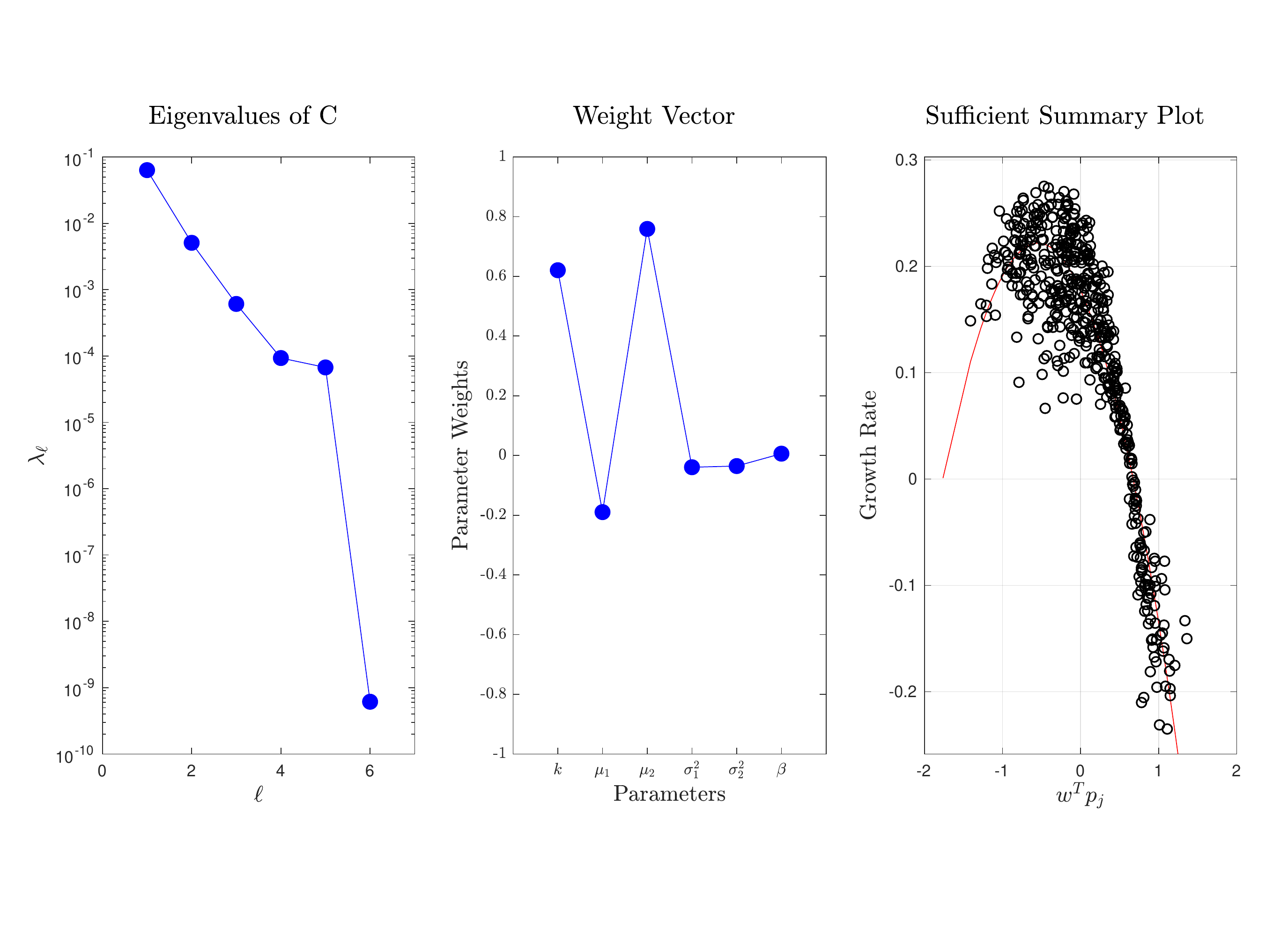}
    \vspace{-0.5in}
    \caption{\footnotesize{\textit{Double Beam Instability:} Global Sensitivity Analysis (25\%) indicating the eigenvalues (left), parameter weights (center), and sufficient summary plot (right) of the one-dimensional active subspace, which represents $\eta_1 = 91.55\%$ of the total information.}} 
    \label{fig:BiMax25}
\end{figure}

The simulations included variations of $1\%$, $5\%$, $15\%$, and $25\%$ of these nominal parameter values, with results presented in Fig.~\ref{fig:BiMax1}-\ref{fig:BiMax25}, and the parameter weights for each of the variational studies are included in Table~\ref{tab:BiMaxDBw}.  As before, figures feature the eigenvalues on the left panel, the parameter weight vector $w$ in the center panel, and the sufficient summary plot of the one-dimensional active subspace in the right panel. 
Throughout each of the simulations, we note that the wavenumber $k$ and mean velocity $\mu_2$ are the most influential on the growth rate. Generally, these parameters are negatively correlated with $\gamma$, which is typically a decreasing function of the active variable $y$. However, the growth of the parabolic component of the reduced approximation indicates that a maximal growth rate is achieved within the parameter regimes corresponding to the $15\%$ and $25\%$ variations. Hence, one can maximize or minimize the rate of instability by altering the physical structure of the beams, in this case the initial frequency of the spatial perturbation $k$ and the mean velocity $\mu_2$. Additionally, notice that the maximal growth rate is approximately equal to that of the Two-Stream instability, and hence these phenomena exhibit similar intensities.

\begin{figure}[t]
\centering
\begin{subfigure}{.5\textwidth}
    \includegraphics[width=0.95\linewidth]{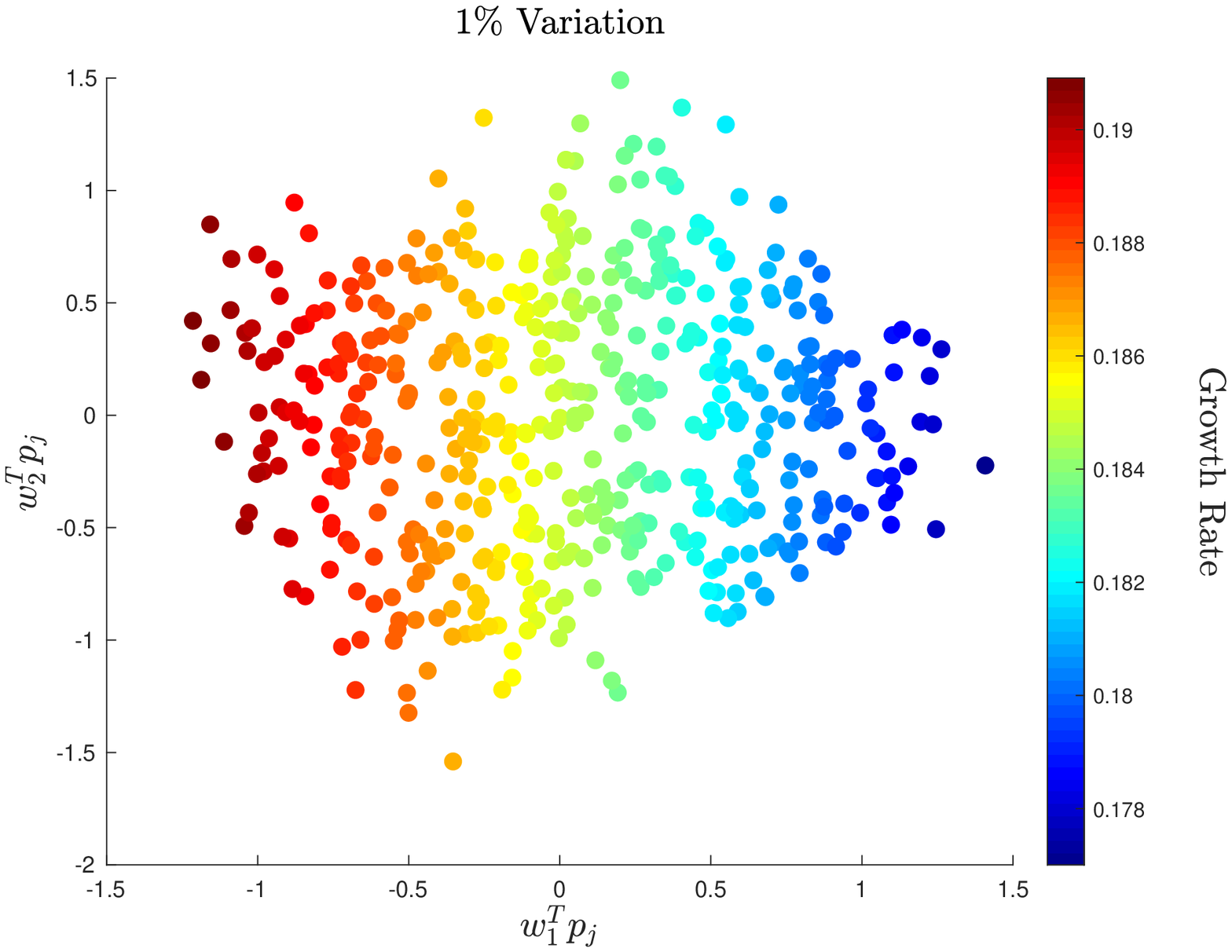}
\end{subfigure}%
\begin{subfigure}{.5\textwidth}
  \includegraphics[width=0.95\linewidth]{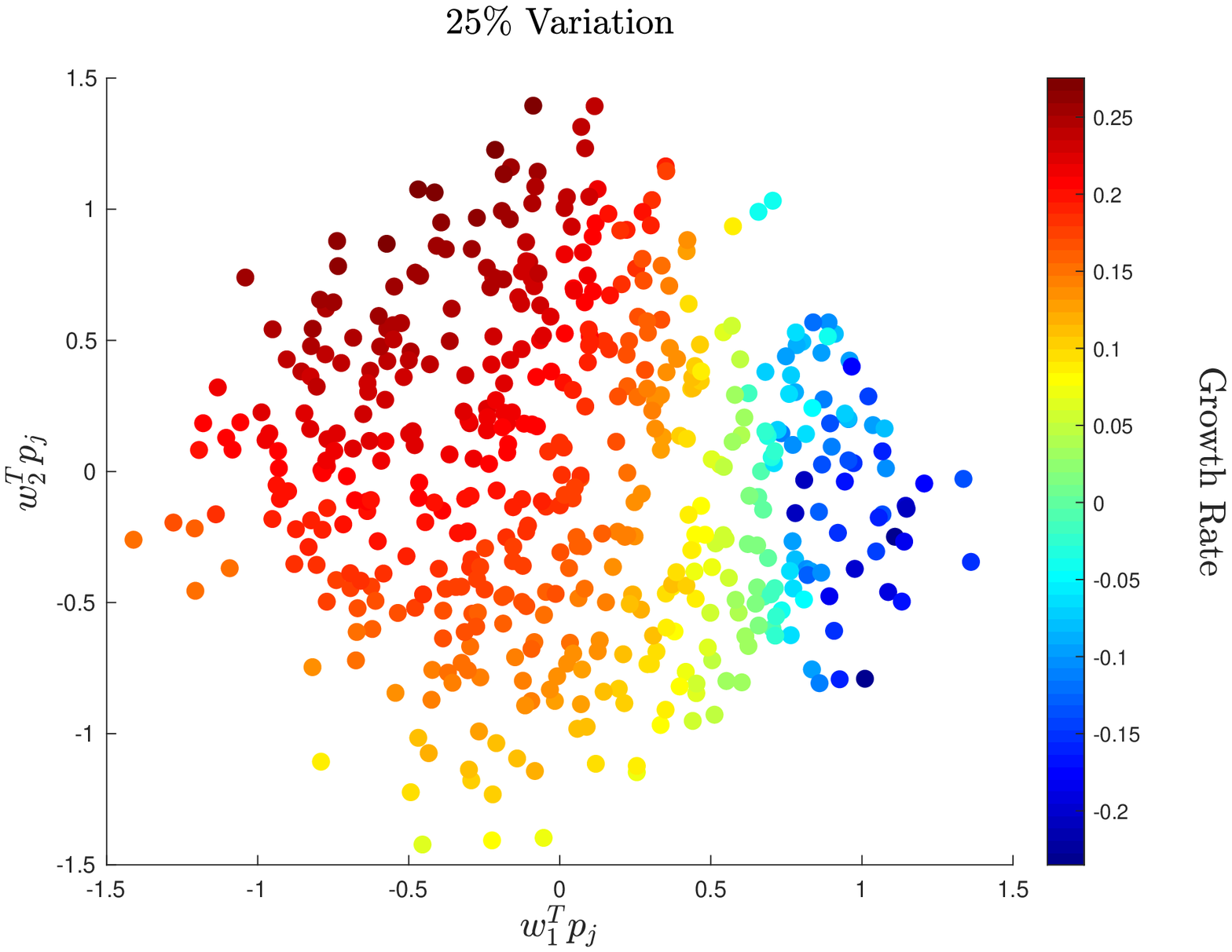}
\end{subfigure}
\caption{\footnotesize{Two-Dimensional Sufficient Summary Plots for the Double Beam Instability: (left) $1\%$ global sensitivity analysis, which represents $\eta_2 = 99.99\%$ of the total variation, and (right) $25\%$ global sensitivity analysis, which represents $\eta_2 = 98.89\%$ of the total variation.
Notice that a one-dimensional representation (see Fig. \ref{fig:BiMax25}) cannot capture the variability expressed in the growth rate as a function of model parameters.}}
\label{fig:BiMaxSSP25}
\end{figure}

The reduced dimension approximation, calculated as 
$$ \eta_1 = \frac{\lambda_1}{\sum_{i=1}^6 \lambda_i},$$ 
is also included in the caption of each figure to demonstrate the percentage of the total variation captured in the one-dimensional subspace. 
More generally, we define 
$$\eta_k = \frac{\sum_{i=1}^k \lambda_i}{\sum_{i=1}^6 \lambda_i}$$ 
for $k = 1, ..., 5$ (and $\eta_6 = 1$) to represent the variation captured by the $k$-dimensional active subspace approximation.
As can be examined by the sufficient summary plot, the one-dimensional active subspace is linear for the 1\%  and 5\% variational studies (Fig.~\ref{fig:BiMax1}-\ref{fig:BiMax5}). For variations of 15\% and greater (Fig.~\ref{fig:BiMax15}-\ref{fig:BiMax25}), the sufficient summary plots become parabolic. This is also observed in Table~\ref{tab:BiMaxDBpf}, which displays a larger magnitude of the quadratic coefficient as the variation increases, e.g. $-2 \times 10^{-4}$ for 1\% variation and $-0.1492$ for 25\% variation, along with the corresponding transition from linear to parabolic curves in Fig.~\ref{fig:DB_Polyfits}.  As expected, the total variation captured by the one-dimensional active subspace drops from $\eta_1 =  99.92\%$ for the 1\% variational study (Fig.~\ref{fig:BiMax1})  to $\eta_1 = 91.55\%$ for the 25\% study (Fig.~\ref{fig:BiMax25}). Although $\eta_1$ does not become as small as that of the Two-Stream $25\%$ variational study, we begin to see in Fig.~\ref{fig:BiMax25} that a one-dimensional active subspace is no longer a strong approximation to the domain of $g(p)$, given the decreasing spectral gap.
In comparison, the growth rate as a function of the first and second parameter weight vectors are plotted for both the 1\% and 25\% simulations in Fig.~\ref{fig:BiMaxSSP25}. The corresponding total approximation of the two-dimensional active subspace is $\eta_2 = 99.995\%$ for the 1\% perturbations and $\eta_2 =98.8\%$ for the 25\% perturbation simulation. Indeed, in the left plot in Fig.~\ref{fig:BiMaxSSP25} representing the 1\% simulation, the growth rate as shown by the vertical banded colors does not change along the vertical axis, which represents the second parameter weight vector. On the other hand, for the 25\% case in the right plot in  Fig.~\ref{fig:BiMaxSSP25}, the growth rate can change significantly along both the horizontal and vertical axes, corresponding to the first and second parameter weight vectors.

As for the Double Beam instability, an explicit approximation of the growth rate can be constructed for each parameter study. For instance, if the normalized parameters are varied by $10\%$ from their baseline value (see Tables \ref{tab:BiMaxDBw} and \ref{tab:BiMaxDBpf}) the growth rate is well-approximated by 
$\gamma_{DB} \approx h(y)$,
where 
$$ y = 0.80 p_1 - 0.14p_2 + 0.58p_3 + 0.05p_4 + 0.05p_5 - 0.0002 p_6$$ 
is the active variable that can be represented in terms of the original variables using
\begin{eqnarray*}
p_1 = 20 (k -0.5), \qquad
p_2 = 10 \mu_1, \qquad
p_3 = 2.5 (\mu_2-4), \qquad\\
p_4 = 20 (\sigma_1^2 -0.5), \qquad
p_4 = 20 (\sigma_2^2 -0.5), \qquad
p_6 = 20 (\beta -0.5),
\end{eqnarray*}
and $h$ is the quadratic function defined by 
$$h(y) = -0.021y^2 - 0.055y + 0.1838.$$
Simplifying $y$ in terms of the original variables yields
$$ y = 16(k-0.5) - 1.4\mu_1 + 1.45(\mu_2 - 4) + (\sigma_1^2 - 0.5) + (\sigma_2^2 - 0.5) - 0.004(\beta - 0.5).$$ 
This expression can then be inserted into $h(y)$ to produce an explicit function for the growth rate in terms of the original parameters, namely
$\gamma_{DB}(k,\mu_1, \mu_2, \sigma_1^2, \sigma_2^2, \beta)$.




\begin{table}[t]
    \centering
    \begin{tabular}{c|c |c|c|c|c|c }
     & \multicolumn{6}{c}{\textbf{Parameter Weights}}  \\
    Variation & \multicolumn{6}{c}{$w^Tp = w_1p_1 + w_2p_2 + w_3p_3  + w_4p_4 + w_5p_5 + w_6p_6$}  \\ 
     \cline{2-7}
     (\%) & $w_1$ & $w_2$ & $w_3$ & $w_4$ & $w_5$ & $w_6$\\
    \hline 
     1 & 0.8666 & -0.1159 & 0.4640 & 0.1006 & 0.1009 & $-4.6515\times 10^{-5}$   \\
     5   &  0.8513 & -0.1235 & 0.4943  &  0.0887 & 0.0888 &  -0.0002 \\
     10 & 0.8017 &  -0.1438 & 0.5752  & 0.0539 & 0.0542 & -0.0002  \\
      15 & 0.6970 &  -0.1739 & 0.6956  &  -0.0077 & -0.0063 & 0.0031  \\
     20 & 0.6423  &  -0.1843  &  0.7421  & -0.0333  & -0.0395  & -0.0065   \\ 
     25 & 0.6206  & -0.1898 &  0.7589 &  -0.0393 &  -0.0354 &  0.0061 \\
    \end{tabular}
    \caption{\footnotesize{Double Beam Instability Parameter Weights}}
    \label{tab:BiMaxDBw}
\end{table}

\begin{table}[t]
\hspace{-0.5in}
\begin{varwidth}[b]{0.42\linewidth}
    \centering
        \begin{tabular}{c|c | c| c}
      &  \multicolumn{3}{c}{\textbf{Polynomial Fit}}  \\ 
    Variation & \multicolumn{3}{c}{$h(w^Tp)=h(y)=a_2y^2+a_1y + a_0$} \\
    \cline{2-4}
     (\%) & $a_2$ & $a_1$ & $a_0$ \\
    \hline 
     1 &  -0.0002 & -0.0053 & 0.1848 \\
      5   &  -0.0046 &  -0.0267 &   0.1845\\
     10 & -0.0213 &  -0.0550 &    0.1838  \\
     15 & -0.0612 &   -0.0886 &    0.1850  \\
     20 & -0.1179 &  -0.1321 &   0.1842 \\
     25 & -0.1492 &  -0.1632  & 0.1772 \\
    \end{tabular}
    \caption{\footnotesize{Double Beam Instability Coefficients}}
    \label{tab:BiMaxDBpf}
\end{varwidth}
    \hspace{0.2in}
\begin{minipage}[b]{0.42\linewidth}
    \includegraphics[width=80mm]{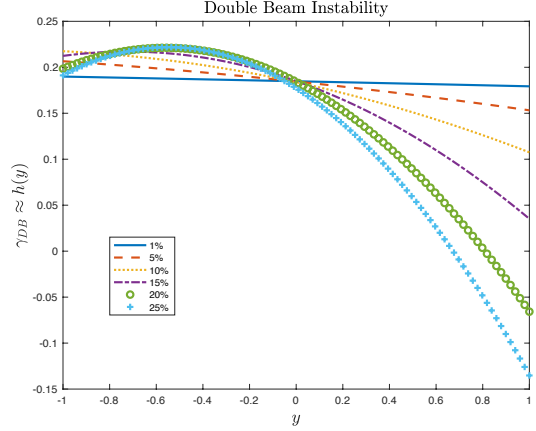}
    \vspace{-0.5in}
    \captionof{figure}{\footnotesize{Double Beam Instability 2nd-order Polynomial Fits}}
    \label{fig:DB_Polyfits}
  \end{minipage}
\end{table}


\vspace{5pt}



\subsubsection{Bump-on-Tail Instability}

In addition to the Double Beam instability of the previous section, we investigate the Bump-on-Tail instability arising from two ionic beams of differing density and mean velocity \cite{Balmforth}. Here, the baseline parameter values are $k=0.5, \mu_1=0, \mu_2=4, \sigma_1^2=0.5, \sigma_2^2=0.5$, and $\beta=0.8$. Similarly, a total number of $N=512$ samples were drawn in the parameter space.

\begin{figure}
    \centering
    \includegraphics[scale=0.5]{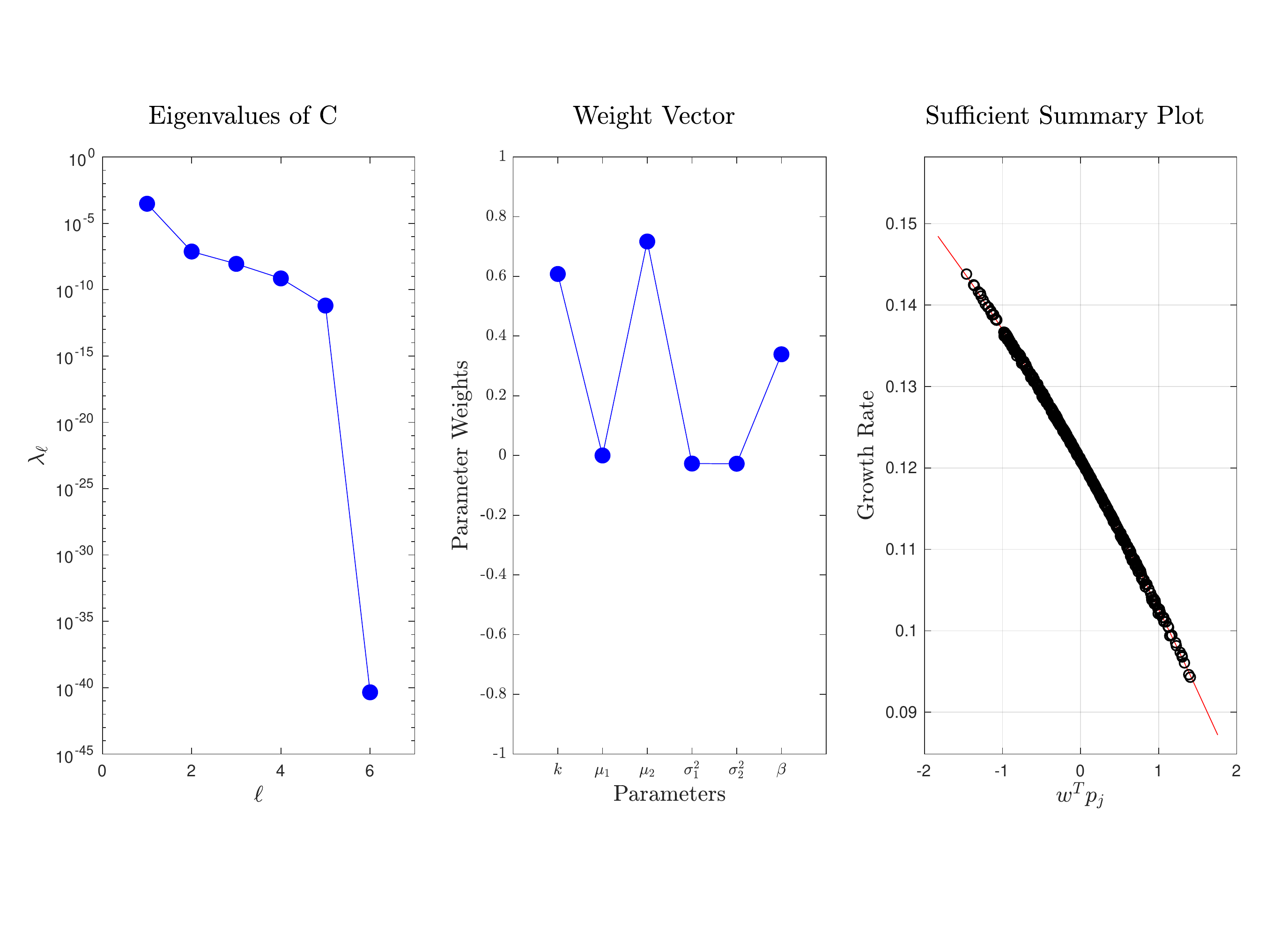}
    \vspace{-0.5in}
    \caption{\footnotesize{\textit{Bump-on-Tail Instability:} Global Sensitivity Analysis (1\%) indicating the eigenvalues (left), parameter weights (center), and sufficient summary plot (right) of the one-dimensional active subspace, which represents $\eta_1 = 99.97\%$ of the total variation.}}
    \label{fig:BoT1}
\end{figure}


\begin{figure}
    \centering
    \includegraphics[scale=0.5]{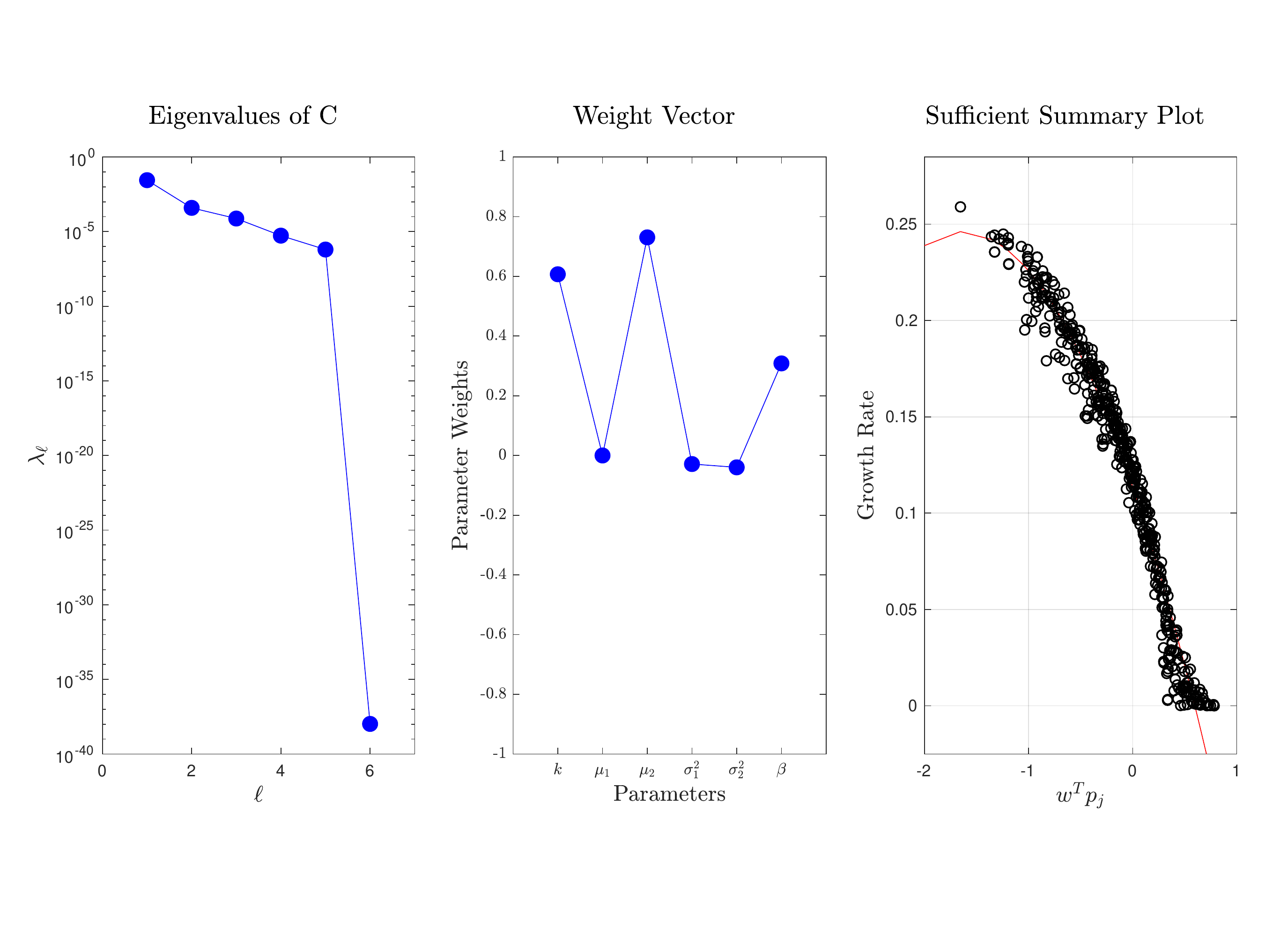}
    \vspace{-0.5in}
    \caption{\footnotesize{\textit{Bump-on-Tail Instability:} Global Sensitivity Analysis (10\%) indicating the eigenvalues (left), parameter weights (center), and sufficient summary plot (right) of the one-dimensional active subspace, which represents $\eta_1 = 98.34\%$ of the total variation.}}
    \label{fig:BoT10}
\end{figure}


\begin{figure}
    \centering
    \includegraphics[scale=0.5]{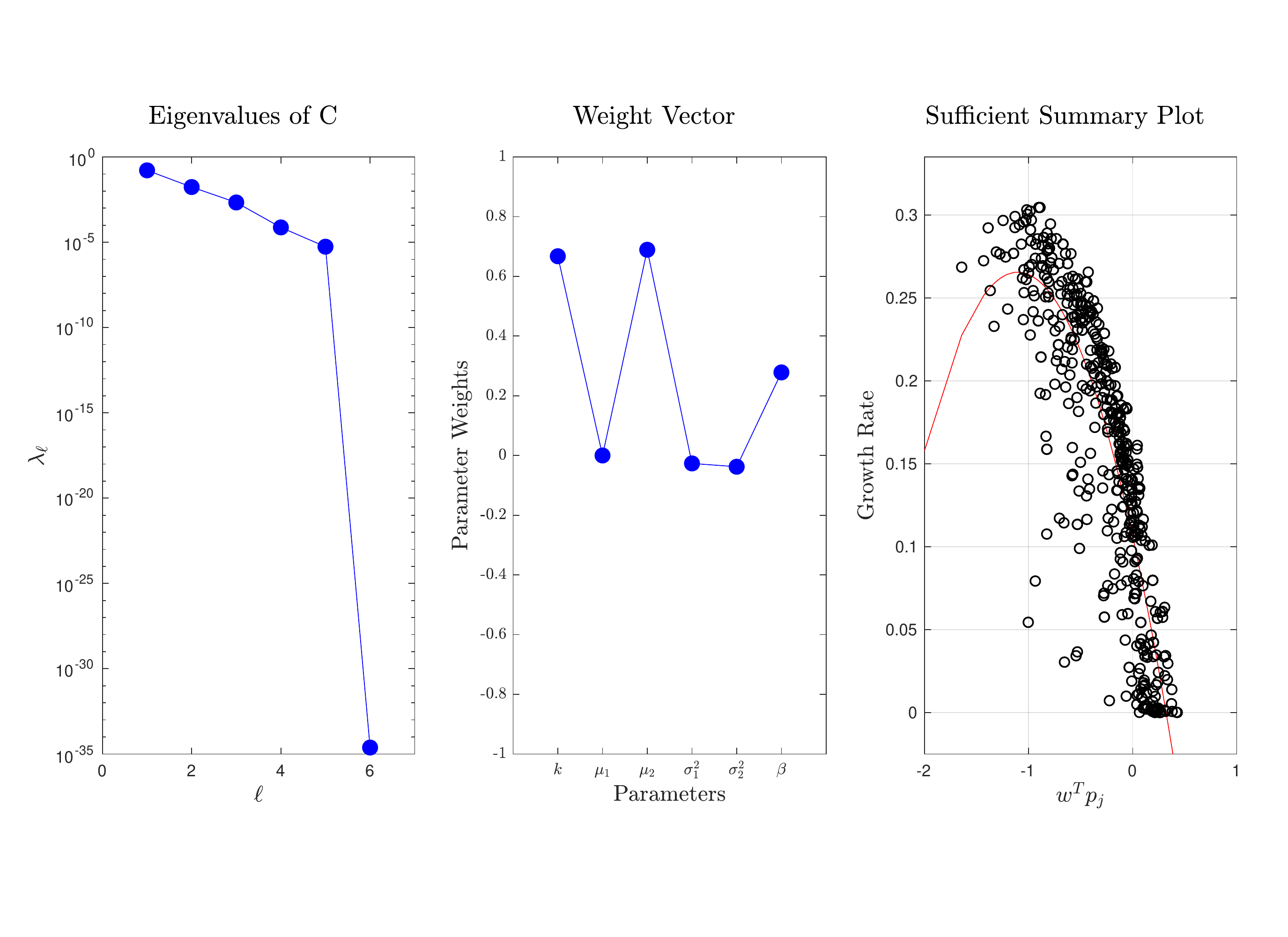}
    \vspace{-0.5in}
    \caption{\footnotesize{\textit{Bump-on-Tail Instability:} Global Sensitivity Analysis (25\%) indicating the eigenvalues (left), parameter weights (center), and sufficient summary plot (right) of the one-dimensional active subspace, which represents $\eta_1 = 89.38\%$ of the total variation.}}
    \label{fig:BoT25}
\end{figure}

The simulations included variations of $1\%$, $10\%$, and $25\%$ of these nominal parameter values, with results presented in Figs.~\ref{fig:BoT1}-\ref{fig:BoT25}, and parameter weights included in Table~\ref{tab:BiMaxBoTw}.  
As shown in the sufficient summary plot, the one-dimensional active subspace is linear for the 1\%  variation study (Fig.~\ref{fig:BoT1}), but becomes parabolic for the 10\% variation study and greater (Fig.~\ref{fig:BoT10}-\ref{fig:BoT25}). This is also observed in Table~\ref{tab:BiMaxBoTpf}, which displays an increasing magnitude of the quadratic coefficient as the variation increases, e.g. $-1.2\times 10^{-3}$ for 1\% variation and $-0.1308$ for 25\% variation, as well as in the polynomial fits in Fig.~\ref{fig:BoT_Polyfits} that gradually become more nonlinear with increased parameter variation.  The information captured by the one-dimensional active subspace drops from $\eta_1 =  99.97\%$ for the 1\% variation study (Fig.~\ref{fig:BoT1})  to $\eta_1 = 89.38\%$ for the 25\% study (Fig.~\ref{fig:BoT25}). In accordance with the reduced spectral gap, we observe that a one-dimensional active subspace is no longer a strong approximation to the domain of $g(p)$, similar to the Double Beam equilibrium 25\% variation study.  
As for the Double Beam instability, the wavenumber $k$ and mean velocity $\mu_2$ are the most influential parameters on the growth rate and generally remain negatively correlated with $\gamma$. However, the density $\beta$ of the larger Maxwellian is also moderately influential within this region of the parameter space; hence, the values of $\gamma$ in the Bump-on-Tail instability will depend on $\beta$ in a non-negligible manner. As before, the influence of the parabolic component of the reduced approximation for the $10\%$ and $25\%$ variations indicates that a maximal growth rate is achieved within these parameter regimes. Thus, the rate of instability can be suitably increased or decreased by altering the initial frequency of the perturbation $k$, the mean velocity $\mu_2$, and the relative density of the larger beam, given by $\beta$. In addition, the maximal value attained by the growth rate is analogous to that of the Two-Stream instability, and hence these phenomena can manifest with similar intensity. In contrast, we note that the volatility (i.e., the spread) in the values of the growth rate for either of the Bi-Maxwellian equilibria is significantly greater than that of the Two-Stream equilibrium.

Again, an explicit approximation of the growth rate can be constructed for each parameter variation study. For instance, if the normalized parameters are varied by $10\%$ from their baseline values (see Tables \ref{tab:BiMaxBoTw} and \ref{tab:BiMaxBoTpf}) the growth rate is well-approximated by 
$\gamma_{BoT} \approx h(y)$,
where
$$ y = 0.61 p_1 + 0.73p_3 - 0.03p_4 - 0.04p_5 + 0.28 p_6$$ 
is the active variable that can be represented in terms of the original variables using
\begin{eqnarray*}
p_1 = 20 (k - 0.5), \qquad
p_2 = 10 \mu_1, \qquad
p_3 = 2.5 (\mu_2-4), \qquad\\
p_4 = 20 (\sigma_1^2 -0.5), \qquad
p_4 = 20 (\sigma_2^2 -0.5), \qquad
p_6 = 12.5 (\beta -0.8),
\end{eqnarray*}
and $h$ is the quadratic function defined by 
$$h(y) = -0.048y^2 - 0.161y + 0.114.$$
As for the other instabilities, these expressions can then be simplified to produce an explicit function for the growth rate $\gamma_{BoT}(k,\mu_1, \mu_2, \sigma_1^2, \sigma_2^2, \beta)$ in terms of the original parameters.

\begin{table}[t]
    \centering
    \begin{tabular}{c|c |c|c|c|c|c }
     & \multicolumn{6}{c}{\textbf{Parameter Weights}}  \\
    Variation & \multicolumn{6}{c}{$w^Tp = w_1p_1 + w_2p_2 + w_3p_3  + w_4p_4 w_5p_5 + w_6p_6$}  \\ 
     \cline{2-7}
     (\%) & $w_1$ & $w_2$ & $w_3$ & $w_4$ & $w_5$ & $w_6$\\
    \hline 
     1 & 0.6081  &   $1.1102\times10^{-16}$ &   0.7168 &  -0.0271 &  -0.0276 &    0.3390 \\
     5 & 0.6019  &  0  &  0.7339 &  -0.0295 & -0.0398 &   0.3109\\
     10 &  0.6071 & 0  &  0.7307 &  -0.0286 &  -0.0397 &   0.3085\\
     15 & 0.6230 & 0  &  0.7077 &  -0.0258 &  -0.0351 &    0.3303 \\
     25 & 0.6675 &  $-1.1102\times10^{-16}$  &  0.6891 &  -0.0266 &  -0.0377  &  0.2783 \\
     \hline
     \text{Global} & 0.5876 &  -0.0579  &  0.2897  &  0.0634 &   0.0431  &  0.7494
    \end{tabular}
    \caption{\footnotesize{Bi-Maxwellian Bump-on-Tail Instability Parameter Weights}}
    \label{tab:BiMaxBoTw}
\end{table}


\subsubsection{Global Bi-Maxwellian Approximation}
\begin{figure}
    \centering
    \includegraphics[scale=0.5]{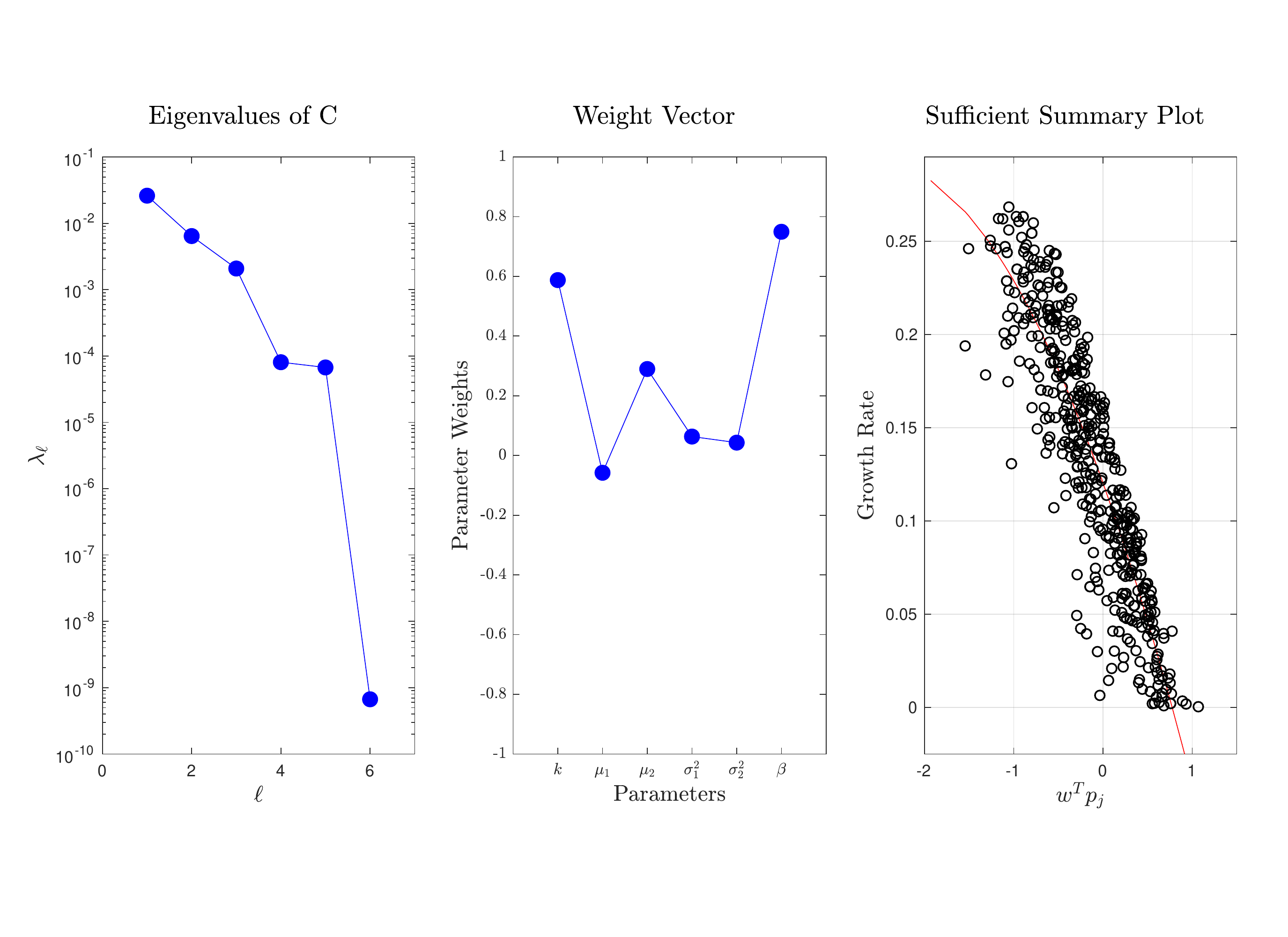}
    \vspace{-0.5in}
    \caption{\footnotesize{\textit{Global Approximation for Bi-Maxwellian Distribution:} Global Sensitivity Analysis indicating the eigenvalues (left), parameter weights (center), and sufficient summary plot (right) of the one-dimensional active subspace, which represents $\eta_1 = 75.09\%$ of the total variation.}}
    \label{fig:Global}
\end{figure}

\begin{figure}[t]
\centering
\begin{subfigure}{.5\textwidth}
    \includegraphics[width=0.95\linewidth]{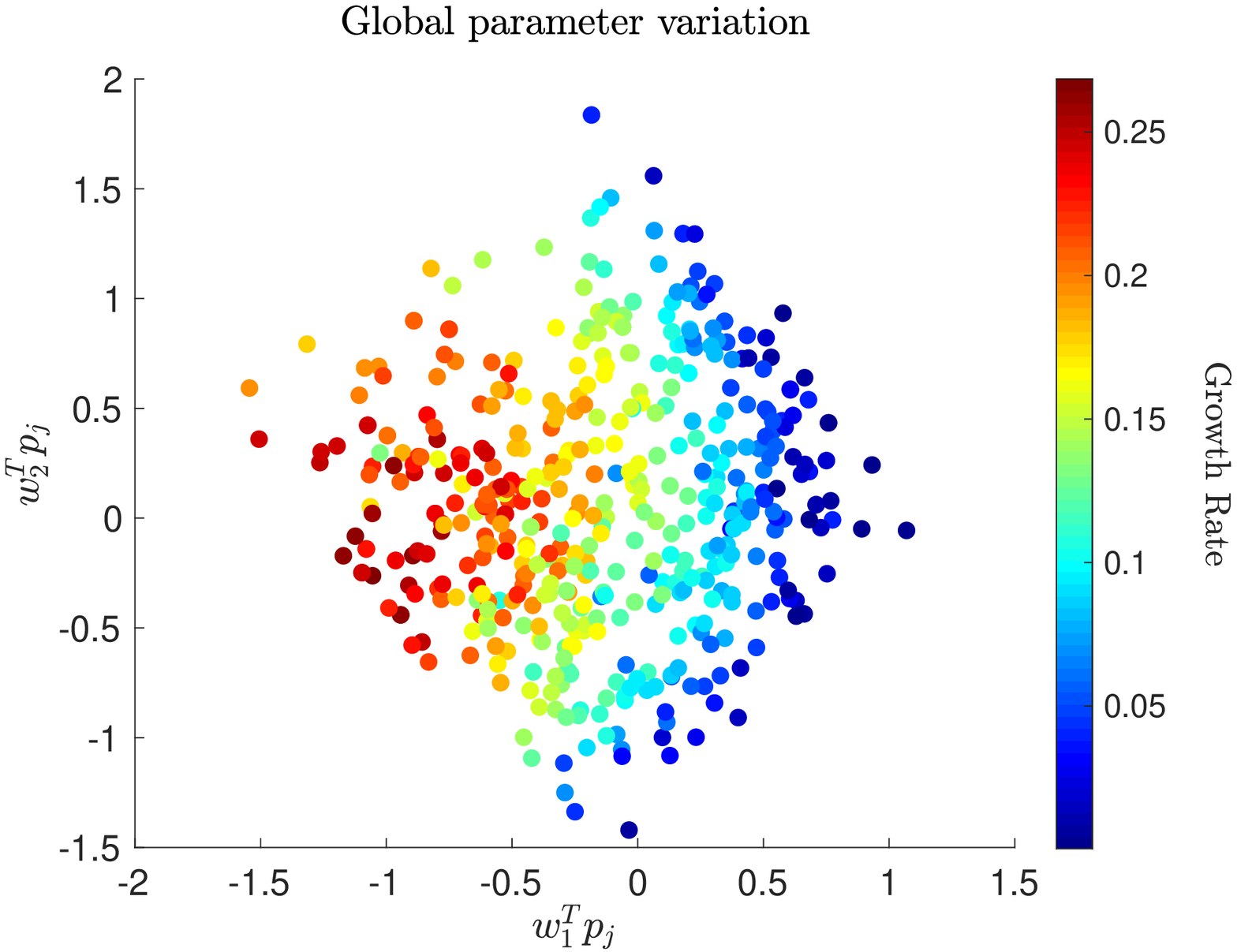}
\end{subfigure}%
\begin{subfigure}{.5\textwidth}
  \includegraphics[width=0.95\linewidth]{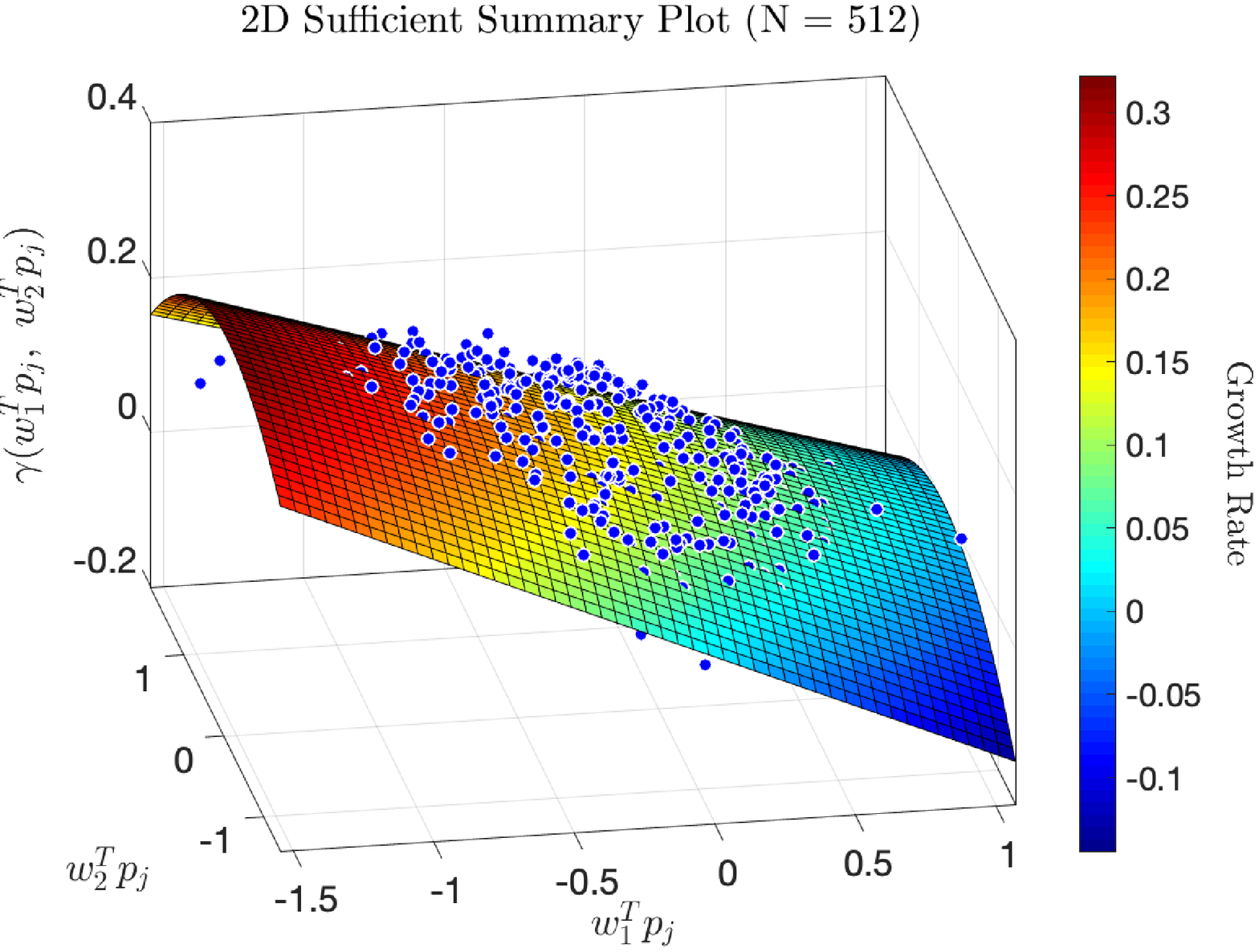}
\end{subfigure}
\caption{\footnotesize{Two-Dimensional Sufficient Summary Plots for the Global approximation of the Bi-Maxwellian Distribution: (left) Two-dimensional scatter plot of the outputs and (right) Three-dimensional plot with best multivariate quadratic approximation,
representing $\eta_2 = 93.58\%$ of the total variation.}}
\label{fig:Global_3D}
\end{figure}

Finally, we also consider an approximation of the Bi-Maxwellian instability rate that is global within the parameter space by using a larger range of parameter values to simultaneously represent both of the previous instabilities.
Rather than perturb the parameters from a set of baseline values, we instead consider a larger hypercube within which they may vary. Here, the lower and upper bound vectors on the parameters $p\propto [k, \mu_1, \mu_2, \sigma_1^2, \sigma_2^2, \beta]$ are given by
$$\ell =  [0.4, -0.1, 3.5, 0.25, 0.25, 0.5]^T \quad 
\mathrm{and} \quad u = [0.6, 0.1, 4.5, 0.75, 0.75, 0.99]^T,$$
respectively.
This implies, for instance, that $\mu_2 = 0.5(p_3+8) \in [3.5, 4.5]$ or equivalently, $p_3 = 2(\mu_2 - 8) \in [-1,1]$.
Results of the one-dimensional active subspace decomposition are presented in Fig.~\ref{fig:Global}. Unfortunately, because of the differing dominant parameter dependencies of the two instabilities, a single active variable is insufficient to capture their distinct behaviors, though the same variables ($k$, $\mu_2$, and $\beta$) remain the most influential.
Thus, a two-dimensional decomposition was also performed (Fig.~\ref{fig:Global_3D}) in order to retain sufficient variation in $\gamma$.
The associated weight vectors are
$$w_1 = [0.588, -0.058, 0.29, 0.06, 0.04, 0.75]^T,$$
which can be seen in Fig. \ref{fig:Global}, 
and
$$w_2 = [-0.567, 0.097, -0.488, 0.143, 0.163, 0.619]^T,$$
representing the two linear combinations of parameters that are most influential on the growth rate.
With this, a global parameter approximation $\gamma_G \approx h(y_1, y_2)$ is obtained using a nonlinear least-squares fit to produce a polynomial of desired degree.
In this way, we have determined the quadratic surface that best fits the growth rate, given by
$$h(y_1,y_2) = 0.1278 - 0.125y_1 + 0.016y_2 + 0.017y_1y_2 - 0.044y_2^2,$$
where
$$y_1 = w_1^Tp \qquad \mathrm{and} \qquad y_2 = w_2^T p.$$
As before, a direct representation of $y_1$ and $y_2$, and hence $\gamma_G$, in terms of the original parameters can be obtained by inverting \eqref{p} for each parameter to find
\begin{eqnarray*}
p_1 = 10 (k -0.5), \qquad
p_2 = 10 \mu_1, \qquad
p_3 = 2 (\mu_2-4), \qquad\\
p_4 = 4 (\sigma_1^2 -0.5), \qquad
p_4 = 4 (\sigma_2^2 -0.5), \qquad
p_6 = 4.08 (\beta -0.745).
\end{eqnarray*}

\begin{table}[t]
\hspace{-0.5in}
\begin{varwidth}[b]{0.42\linewidth}
    \centering
  \begin{tabular}{c|c | c| c}
      &  \multicolumn{3}{c}{\textbf{Polynomial Fit}}  \\ 
    Variation & \multicolumn{3}{c}{$h(w^Tp)=h(y)=a_2y^2+a_1y + a_0$} \\
    \cline{2-4}
     (\%) & $a_2$ & $a_1$ & $a_0$ \\
    \hline 
     1 &  -0.0012  &  -0.0172  &   0.1210 \\
     5 & -0.0226  &   -0.0902  &   0.1185 \\
     10 & -0.0489 &  -0.1610  &  0.1136 \\
     15 & -0.0825 & -0.2200  &  0.1109 \\
     25 & -0.1308 &   -0.2888 &    0.1061 \\
     \hline
     \text{Global} & -0.0263 &  -0.1348 &   0.1204
    \end{tabular}
    \caption{\footnotesize{Bump-on-Tail Instability Coefficients}}
    \label{tab:BiMaxBoTpf}
\end{varwidth}
    \hspace{0.2in}
\begin{minipage}[b]{0.42\linewidth}
    \includegraphics[width=85mm]{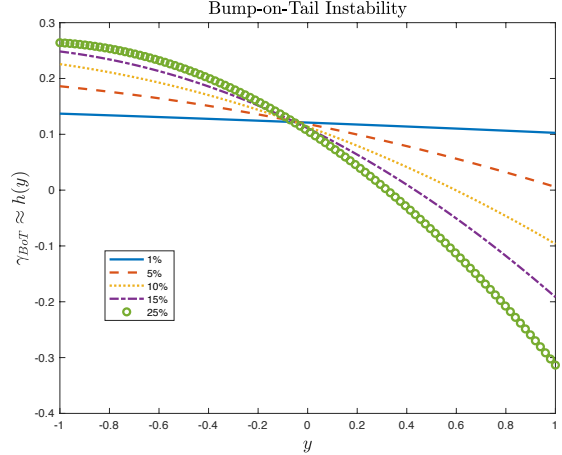}
    \vspace{-0.5in}
    \captionof{figure}{\footnotesize{Bump-on-Tail Instability 2nd-order Polynomial Fits}}
    \label{fig:BoT_Polyfits}
  \end{minipage}
\end{table}


Using the eigenvalues to compute the information contained in this approximation, we find
$$\eta_1 = \frac{\lambda_1}{\sum_{i=1}^6 \lambda_i} = 0.7509 \qquad \mathrm{and} \qquad \eta_2 = \frac{\lambda_1 + \lambda_2}{\sum_{i=1}^6 \lambda_i} = 0.9358.$$
Thus, researchers that possess less precise knowledge concerning the range of parameter values within an experiment can utilize this two-dimensional global parameter approximation to capture more than $93\%$ of the variation in $\gamma$.
Additionally, this approximation shows that while fluctuations in the beam density $\beta$ of the larger Maxwellian give rise to the greatest changes throughout the entire parameter space, this effect is attenuated within the two featured regimes (Double Beam and Bump-on-Tail).

\section{Conclusions}
The formulation and implementation of the active subspace method has allowed us to draw a variety of conclusions. First, the parameters of greatest importance to the rate of instability were determined for each equilibrium. In particular, for both the Double Beam and Bump-on-Tail instabilities, the wavenumber $k$ and mean velocity $\mu_2$ are most influential to the growth rate $\gamma$. This, along with the negative correlation of these parameters with $\gamma$, indicates that the rate of instability can be controlled, and hence increased if desired, by decreasing the spatial frequency of the initial perturbation or the mean velocity of the second beam. Conversely, the instability can be tamed by utilizing a high frequency spatial perturbation or suitably increasing this mean velocity.  

Second, the extremal values of the growth rate were computed on these intervals, and they demonstrate that the intensity of the Bump-on-Tail or Double Beam instability is similar to that of the Two-Stream instability, as the attained growth rates are all around $0.3$ in this dimensionless system. Hence, previous indications \cite{Tokluoglu} that the latter instability must occur with greater intensity or arise on faster timescales may not hold when fluctuations occur in the mean velocities or density of the particle beams, or when differing wavenumbers are considered.
That being said, it is clear from our simulations that the instabilities arising from the Bi-Maxwellian distribution possess greater variability in their rates of growth than does the Two-Stream Instability.

Finally, using the dispersion relation, we were able to construct an analytic representation of the growth rate as a function of model parameters for both the Two-Stream and Bi-Maxwellian equilibria. The dispersion-based growth rate solver enabled accurate, efficient, and inexpensive simulations even for large parameter variations, such as the $20\%$, 
$25\%$, and $50\%$ simulation cases. 
Though the one-dimensional approximations begin to break down as the variation percentage is increased beyond $25-50\%$, two-dimensional active subspace approximations were able to retain the overwhelming majority of information contained in the growth rate.
In particular, simulations of the Two-Stream instability provided the knowledge that $\gamma$ is essentially independent of the mean velocity $\mu$ and allowed for an inexpensive, reduced decomposition with high accuracy.
Additionally, simulations of the Double Beam and Bump-on-Tail instabilities demonstrated the physical differences arising from changes in parameter values.
In either case, sufficiently accurate one-dimensional and two-dimensional active subspace approximations of the growth rate $\gamma$ were acquired.
Finally, a global approximation of the growth rate was obtained for a larger range of parameter values to show that a two-dimensional active subspace representation can describe the behavior of both extreme cases near the Bi-Maxwellian equilibrium.
For all equilibria these approximations yield explicit formulas for the growth rate as a function of system parameters.

\appendix
\section{Dimensionless Model}

\label{sec:AppA}
In order to construct the dimensionless equations~\eqref{VPnd}, we begin with the original Vlasov-Poisson system

\begin{subequations}
\label{VP}
\begin{align}
\partial_{\tilde{t}} \tilde{f} + \tilde{v} \partial_{\tilde{x}} \tilde{f} - \frac{q}{m} \tilde{E} \partial_{\tilde{v}} \tilde{f} = 0\\
\partial_{\tilde{x}} \tilde{E} = \frac{q}{m\epsilon_0} (\tilde{\rho}_0 - \tilde{\rho}(\tilde{t},\tilde{x}))\\
\tilde{\rho}(\tilde{t},\tilde{x}) = m  \int \tilde{f}\ d\tilde{v}.\end{align}
\end{subequations}
Here, $\tilde{f}(\tilde{t},\tilde{x},\tilde{v})$ represents the electron distribution, $\tilde{\rho}(\tilde{t},\tilde{x})$ is the corresponding charge density, and $\tilde{E}(\tilde{t},\tilde{x})$ is the electric field induced by the charge in the system.  Additionally, 
$$\tilde{\rho}_0 = \frac{1}{L_0} \int_0^{L_0} \tilde{\rho}(0,\tilde{x}) \ d\tilde{x}$$ represents a background neutralizing density, and the parameters $q, m$, and $\epsilon_0$ are the charge and mass of a particle and the permittivity of free space, respectively.

For each independent and dependent variable, we introduce a scaling factor:
$$\begin{gathered}
\tilde{t} = T_0t, \quad \tilde{x} = L_0x, \quad \tilde{v} = \frac{L_0}{T_0} v\\
\tilde{f} = F_0f, \quad \tilde{E} = E_0 E, \quad \tilde{\rho} = \tilde{\rho}_0 \rho. 
\end{gathered}$$

Rewriting \eqref{VP} in terms of the scaled variables ultimately yields the system
$$\begin{gathered}
\partial_t f + v \partial_x f - \left (\frac{qE_0T_0^2}{mL_0} \right ) E \partial_v f = 0\\
\partial_x E = \left (\frac{qL_0\tilde{\rho}_0}{\epsilon_0 E_0} \right ) (1-\rho )\\
\rho = \left (\frac{F_0L_0}{\tilde{\rho}_0T_0} \right ) \int f(t,x,v) \ dv.
\end{gathered}$$

In order to reduce the dimension of the parameter space, we normalize the constant quantities within the parentheses.
Doing so implies choosing
\begin{equation}
E_0 = \frac{qL_0\tilde{\rho}_0}{\epsilon_0}, \quad
T_0 = \sqrt{\frac{\epsilon_0 m}{q^2 \tilde{\rho}_0}}, \quad
F_0 = \frac{\sqrt{\tilde{\rho}_0 \epsilon_0 m}}{qL_0}.
\end{equation}
With this scaling, we arrive at 
the dimensionless Vlasov-Poisson system, namely~\eqref{VPnd}.

We assume that the background density, $\tilde{\rho}_0$, and the length scale, $L_0$, have been selected. The remaining parameters, $E_0$, $T_0$, and $F_0$, are then uniquely determined, and \eqref{VPnd} results for any choice of $\tilde{\rho}_0$ and $L_0$.
In particular, a standard choice is to select the length scale of the Debye length, $L_0 =  \lambda_D$, where
$$\lambda_D = \sqrt{\frac{\epsilon_0k_BT}{\tilde{\rho}_0 q^2}},$$
$k_B$ is the Boltzmann constant and $T$ is the temperature, and this yields the time scale of the inverse plasma frequency, namely
$T_0 = \omega_p^{-1}$, 
where
$$\omega_p = \sqrt{\frac{q^2 \tilde{\rho}_0}{\epsilon_0 m}}.$$
These length and time scalings then imply a specific velocity scaling
$$ \tilde{v} = \lambda_D \omega_p v = \sqrt{\frac{k_B T}{m}} v$$
that can influence the value of dimensionless parameters in the equilibrium distribution functions.
For instance, if the temperature $T$ is increased by a certain factor $A>0$, then the dimensionless parameter $\sigma^2$ is also increased by the same factor of $A$.

\section{Dispersion Relations for Two-Stream and Bi-Maxwellian Equilibria}
\label{sec:AppB}

For the equilibria under consideration, explicit formulas for the growth rate within the linear regime can be computed in terms of the $Z$-function.  In particular, assuming that the charge distribution has the form
$$f(t, x, v) = f_{eq}(v)(1 + \delta f(t, x, v))$$
where
$\delta f(t,x,v) =  \alpha\exp\left (i[kx - \omega t]\right)$,
one can express the growth rate $\gamma$ as a function of the frequency $\omega$ for fixed wavenumber $k$ and certain spatially-homogeneous equilibria.  
The plasma dispersion relation, obtained by linearizing \eqref{VP} about $f_{eq}$ and searching for plane wave solutions, is
\begin{equation}
\label{dispApp}
\eps(k, \omega) = 1 - \frac{1}{k^2} \int_{-\infty}^\infty \frac{f'_{eq}(v)}{v - \omega/k} dv.
\end{equation}

Prior to computing the dispersion relation for the equilibria of interest, we will perform this calculation for a Maxwellian distribution and then use this information to generalize the expression for the Two-Stream and Bi-Maxwellian cases.
So, first consider a single Maxwellian equilibrium, for which the distribution of particle velocities is given by
$$f_{M}(v; \mu, \sigma^2) = \frac{1}{\sqrt{2\pi \sigma^2}} \exp \left (-\frac{1}{2\sigma^2} \vert v -\mu \vert^2 \right ).$$ 
Taking a derivative of this function, writing $u = \omega/k$, and substituting $z = \frac{1}{\sqrt{2\sigma^2}}(v-\mu)$ within the resulting integral of \eqref{dispApp} yields
$$\eps(k, ku) = 1 + \frac{1}{\sqrt{\pi} \sigma^2 k^2} \int_{-\infty}^\infty \frac{z}{z - \frac{1}{\sqrt{2\sigma^2}} (u-\mu)} e^{-z^2} \ dz.$$
Next, we let 
$$A(u) = \frac{1}{\sqrt{2\sigma^2}} ( u - \mu)$$
and rewrite the integrand as
$$\frac{z}{z - A(u)} e^{-z^2}  = e^{-z^2} + \frac{A(u)}{z - A(u)} e^{-z^2}.$$
Then, the integral of the first term can be computed explicitly as $\sqrt{\pi}$, while the second involves the $Z$-function given by \eqref{Z}.
Ultimately, we find
\begin{equation}
\label{epsMax}
\eps(k, ku) = 1 + \frac{1}{\sigma^2 k^2} \left [ 1 + A(u) Z(A(u)) \right ],
\end{equation}
the roots of which yield an implicit representation for $\gamma =  \text{Im}(ku)$.

For the Two-Stream equilibrium, the distribution of particle velocities is given by
\begin{equation}
f_{TS}(v; \mu, \sigma^2) =\frac{1}{\sqrt{2\pi \sigma^2}} |v-\mu|^2 \exp \left (-\frac{1}{2\sigma^2} | v -\mu |^2 \right).
\end{equation}
Similar to the Maxwellian distribution, we first differentiate to find
$$ f'_{eq}(v) = \frac{2}{\sqrt{2\pi \sigma^2}} (v - \mu) \exp \left (-\frac{1}{2\sigma^2} | v -\mu |^2 \right )-\frac{1}{\sigma^2\sqrt{2\pi \sigma^2}} (v - \mu)^3 \exp \left (-\frac{1}{2\sigma^2} | v -\mu |^2 \right ),$$
which we decompose as 
\[f'_{eq}(v) = g_1(v) - g_2(v). \]
As the dispersion relation is linear in the contribution of $f_{eq}'(v)$, we use \eqref{dispApp} to find
$$\eps(k,\omega) = 1 - \frac{1}{k^2} \int \frac{g_1(v)}{v-\omega/k} dv + \frac{1}{k^2}\int \frac{g_2(v)}{v-\omega/k} dv. $$
Proceeding with the $g_1$ integrand and using similar substitution techniques as for the Maxwellian, including the change of variables $t = \frac{1}{\sqrt{2\sigma^2}} ( v -\mu )$, we have
\begin{align*}
    \int \frac{g_1(v)}{v-\frac{\omega}{k}} dv & = \frac{2}{\sqrt{2\pi \sigma^2}} \int \frac{(v - \mu) \exp \left (-\frac{1}{2\sigma^2} | v -\mu |^2 \right )}{v-\frac{\omega}{k}} dv , \\
    & = \frac{2}{\sqrt{\pi}} \int \frac{t e^{-t^2}}{t-A(u)}dt , \\
 & = \frac{2}{\sqrt{\pi}} \left[\int e^{-t^2} dt + A(u)\int  \frac{e^{-t^2}}{t-A(u)}dt\right] , \\
 & = {2} \left[1 + A(u)Z(A(u))\right],
\end{align*}
where $u=\omega/k$ and $A(u)= \frac{1}{\sqrt{2\sigma^2}}\left(u-\mu \right)$. 
To compute the second integral, we perform the same change of variables to find 
 \begin{align*}
     \int \frac{g_2(v)}{v-\frac{\omega}{k}} dv & = \frac{1}{\sigma^2\sqrt{2\pi \sigma^2}}  \int \frac{(v - \mu)^3 \exp \left (-\frac{1}{2\sigma^2} | v -\mu |^2 \right )}{v-\frac{\omega}{k}} dv, \\
     & = \frac{2}{\sqrt{\pi}}   \int \frac{t^3 e^{-t^2}}{t - A(u)} dt  
 \end{align*}
where $u=\omega/k$ and $A(u)= \frac{1}{\sqrt{2\sigma^2}}\left(u-\mu \right)$. 
Next, we make repeated use of the identity
$$ \frac{t}{t-A(u)} = 1+ \frac{A(u)}{t-A(u)}$$
in order to simplify the rational portion of the integrand, resulting in
$$\frac{t^3}{t-A(u)} = t^2 + tA(u) + A(u)^2 +\frac{A(u)^3}{t-A(u)}.$$
With this, we can perform the integration in each portion of the integral, which yields
\begin{align*}
    \int \frac{g_2(v)}{v-\frac{\omega}{k}} dv & = \frac{2}{\sqrt{\pi}}   \int \left[t^2 + tA(u) + A(u)^2 +\frac{A(u)^3}{t-A(u)}\right] e^{-t^2} dt \\
    & = \frac{2}{\sqrt{\pi}}  \left[ \int t^2 e^{-t^2} dt +  A(u)\int t e^{-t^2}dt + A(u)^2\int e^{-t^2} dt +A(u)^3 \int \frac{e^{-t^2}}{t-A(u)}dt \right] \\
    & = \frac{2}{\sqrt{\pi}}  \left[\frac{\sqrt{\pi}}{2} + A(u)^2\sqrt{\pi}+A(u)^3 \sqrt{\pi}Z(A(u))\right] \\
    & = 1 + 2A(u)^2 +2A(u)^3 Z(A(u)). 
\end{align*}
Reassembling the results of the $g_1$ and $g_2$ terms, we finally arrive at a simplified expression for the dispersion relation, namely
\begin{equation}
\eps_{TS} (k,ku) =1 - \frac{1}{k^2} \left[1 - 2A(u)^2 +2\left(A(u)-A(u)^3\right)Z(A(u))\right],
\end{equation}
the roots of which yield an implicit representation for $\gamma_{TS}(k,\mu,\sigma^2) =  \text{Im}(ku)$.

Next, we return to \eqref{dispApp} and simplify the dispersion relation for the Bi-Maxwellian equilibrium, for which the distribution of particle velocities is given by
\begin{equation}
    f_{BM}(v) =\frac{\beta}{\sqrt{2\pi \sigma_1^2}} \exp \left (-\frac{1}{2\sigma_1^2} |v -\mu_1|^2 \right ) + \frac{1-\beta}{\sqrt{2\pi \sigma_2^2}} \exp \left (-\frac{1}{2\sigma_2^2} |v -\mu_2|^2 \right ),
\end{equation}
 where $0 < \beta < 1$. 
 Noting that this distribution is merely a convex combination of two Maxwellians, we use the linearity of the dispersion relation and the previously-computed formula \eqref{epsMax} for each Maxwellian separately to find
$$\int \frac{f_{eq}'(v)}{v-\omega/k} dv = -\frac{\beta}{\sigma_1^2} \left[1 + A_1(u)Z(A_1(u))\right] -\frac{1-\beta}{\sigma_2^2} \left[1 + A_2(u)Z(A_2(u))\right],$$
where $u = \omega/k$,
$$A_1(u)=\frac{1}{\sqrt{2\sigma_1^2}}(u-\mu_1) \qquad \mathrm{and} \qquad  A_2(u)=\frac{1}{\sqrt{2\sigma_2^2}}(u-\mu_2).$$ 
Thus, we have a representation for the dispersion relation as
$$ \eps_{BM}(k,ku) = 1 +\frac{\beta}{\sigma_1^2k^2} \left[1 + A_1(u)Z(A_1(u))\right] +\frac{1-\beta}{\sigma_2^2k^2} \left[1 + A_2(u)Z(A_2(u))\right],$$
and the associated growth rate $\gamma_{BM}(k, \mu_1, \mu_2, \sigma_1^2, \sigma_2^2, \beta) = \text{Im}(ku)$ is computed as the root of $\eps_{BM}(k,ku)$ given the wavenumber $k$.


\bibliography{SP_ASI.bib}

\end{document}